\documentclass[lettersize,journal]{IEEEtran}

\usepackage{amsmath,amsfonts}
\usepackage{array}
\usepackage[caption=false,font=normalsize,labelfont=sf,textfont=sf]{subfig}
\usepackage{textcomp}
\usepackage{stfloats}
\usepackage{url}
\usepackage{verbatim}
\usepackage{graphicx}
\usepackage{cite}
\usepackage[english]{babel}
\usepackage{amsmath}
\usepackage{amsfonts}
\usepackage{graphicx}
\usepackage{float} 
\usepackage[colorlinks=true, allcolors=blue]{hyperref}
\usepackage[utf8]{inputenc}
\usepackage{bbm}
\usepackage[ruled,vlined]{algorithm2e}
\usepackage{caption}
\usepackage{booktabs}
\usepackage{multirow}
\usepackage{colortbl}
\usepackage{xcolor}
\usepackage{float} 
\usepackage{booktabs}  
\usepackage{multirow}  
\usepackage{tabularx}  
\usepackage{makecell} 
\usepackage{tabularx} 
\usepackage{arydshln} 
\usepackage{siunitx}
\usepackage{amssymb}

\usepackage{subcaption}

\title{Evaluating the Robustness of Adversarial Defenses in Malware Detection Systems}

\author{
    Mostafa Jafari \qquad Alireza Shameli-Sendi\\
    \IEEEauthorblockA{
        Faculty of Computer Science and Engineering, Shahid Beheshti University (SBU), Tehran, Iran\\
        Email: most.jafari@mail.sbu.ac.ir,\, a\_shameli@sbu.ac.ir
    }
}

\begin{document}
\maketitle

\begin{abstract}
Machine learning-based malware detectors effectively identify malicious Android applications, but remain highly vulnerable to evasion attacks, where minor binary feature perturbations can evade detection. Despite ongoing progress, the lack of comprehensive evaluation frameworks limits a clear understanding of defense robustness.
This study introduces two contributions for adversarial evaluation in binary-constrained malware detection. First, we propose Prioritized Binary Rounding, an efficient technique that converts continuous perturbations into binary features while preserving a high attack success rate and minimal modification budgets. Second, we present the $\sigma$-binary attack, a gradient-based method tailored for binary domains, capable of producing sparse yet highly effective perturbations.
Experimental results in the MalScan dataset show that $\sigma$-binary achieves consistently superior and stable performance. It surpasses state-of-the-art binary-domain approaches against both robust and non-robust defenses.
Advanced defenses such as KDE, DLA, DNN$^{+}$, and ICNN, when evaluated under the proposed $\sigma$-binary attack, exhibit over 90\% attack success with fewer than 10 modified features and reach 100\% under a 20-feature budget. Even PAD-SMA, previously reported to retain 83\% accuracy, is compromised by $\sigma$-binary with 36.3\% success under 10-features and 94.6\% with no feature-budget constraints.
These results highlight the need for precise evaluation tools such as the proposed $\sigma$-binary attack to uncover latent weaknesses and guide the design of truly robust malware detectors.
\end{abstract}


\section{Introduction}

The surge in mobile device attacks has escalated into a critical global concern. In 2023 alone, such attacks increased by 50\% compared to the previous year. According to Kaspersky, their systems blocked 33.8 million instances of malware, adware, and riskware targeting mobile devices, with Android remaining the primary focus of cybercriminals~\cite{kaspersky2024}. This significant increase in Android malware underscores the urgent need for advanced detection strategies to counter increasingly sophisticated threats.

Machine learning (ML) has become a cornerstone of Android malware detection, significantly improving scalability and detection precision \cite{gao2024comprehensive, liu2024unraveling, arp2014drebin, he2022msdroid, liu2022deep}. Among ML-based approaches, static analysis methods, such as Drebin \cite{arp2014drebin} and MaMaDroid~\cite{mariconti2016mamadroid}, have demonstrated notable efficiency. These methods analyze application features without requiring execution, enabling rapid and scalable security assessments. Consequently, static analysis has become a foundational component of modern malware detection pipelines.

Despite these advancements, ML-based malware detectors remain inherently vulnerable to adversarial attacks. These attacks exploit weaknesses in detection models by introducing carefully crafted perturbations to evade detection. Such attacks, which undermine reliability, are broadly categorized into evasion and poisoning attacks. Evasion attacks manipulate test examples during the deployment phase to bypass detection \mbox{\cite{grosse2017adversarial, huang2018adversarial, li2023black, bostani2024evadedroid, hu2022generating}}, whereas poisoning attacks compromise training datasets to degrade detector performance \cite{chen2018automated, bala2022droidenemy}. This study focuses on evasion attacks because of their immediate implications for operational systems.

Evasion attacks can be further divided into problem-space and feature-space attacks. Problem-space attacks modify malware applications directly, such as injecting benign "gadgets" or altering application components to obscure malicious intent \cite{pierazzi2020intriguing, chen2019android, tang2024dapadv}. In contrast, feature-space attacks modify feature vectors extracted from malware while preserving the functionality of the original sample \cite{huang2018adversarial, grosse2017adversarial}. This study focuses exclusively on feature-space attacks due to their controlled evaluation capabilities and practical relevance.

Defending against adversarial attacks remains a significant challenge. Adversarial training, which incorporates adversarial samples into training datasets, has shown promise in improving robustness \cite{grosse2017adversarial}. However, these methods often fail to generalize to novel attack strategies, impose high computational costs, and risk overfitting to specific attacks \cite{carlini2019evaluating}. Other defense mechanisms such as ensemble methods \cite{smutz2016tree, li2020adversarial, ficco2021malware}, auxiliary models such as Variational Autoencoders (VAEs) \cite{li2021robust}, and hybrid frameworks such as PAD-SMA \cite{li2023pad} offer varying degrees of effectiveness. However, these approaches frequently involve trade-offs, including increased complexity and reduced detection accuracy.

Evaluating the adversarial robustness of malware defenses in binary spaces presents significant challenges. Existing approaches face two major limitations. First, gradient-based attacks originally designed for continuous feature spaces are frequently adapted to binary domains using naive binarization techniques, such as deterministic or randomized rounding~\cite{huang2018adversarial}. These adaptations substantially reduce attack effectiveness and fail to accurately capture the discrete nature of binary features. Furthermore, the absence of gradient-based attack formulations explicitly designed for binary spaces stems from the inherent difficulty of optimizing within a non-convex and non-differentiable constraint space. Second, many existing evaluation methodologies rely on generalized attack settings that employ low-iteration gradient-based attacks with fixed hyperparameters across all defense models~\cite{li2023pad, li2021robust, zhou2023malpurifier, li2020adversarial, li2021framework}. This lack of adaptive optimization prevents such evaluations from effectively exposing model-specific vulnerabilities and leads to an overestimation of robustness. Notably, previous studies have reported that black-box attacks, such as Mimicry, can sometimes outperform gradient-based approaches in binary domains~\cite{li2020adversarial, li2023pad} despite having no access to model parameters or architecture. This counterintuitive result arises because gradient-based methods, although privileged with full model knowledge, struggle to optimize effectively under discrete and non-differentiable binary constraints, leading to suboptimal perturbations and reduced attack success.

To address these limitations, this paper introduces a unified and systematic framework for assessing and enhancing model robustness in binary-constrained malware detection. The proposed framework establishes a principled, binary-compliant evaluation process capable of generating functionally valid adversarial examples and providing a reliable foundation for comparative robustness analysis.

In this context, our study makes the following key contributions:

\begin{enumerate}
    \item \textbf{Prioritized Binary Rounding (PBR):} We develop a rounding mechanism that effectively transforms continuous perturbations into valid binary feature representations while preserving high attack success rates and minimizing modification budgets.

    \item \textbf{$\sigma$-binary Attack:} We introduce a novel gradient-based adversarial attack specifically tailored for binary domains, capable of generating sparse yet highly effective adversarial examples through a differentiable optimization process.

    \item \textbf{Comprehensive Experimental Evaluation:} Extensive experiments on the MalScan dataset demonstrate that $\sigma$-binary achieves consistently superior and stable performance compared to state-of-the-art approaches in the binary domain, against both robust and non-robust defense conditions.
\end{enumerate}

Overall, this work establishes a principled and reproducible foundation for adversarial robustness evaluation in malware detection, bridging the gap between continuous-space optimization methods and discrete, real-world feature constraints.

The source code for this study is publicly available at \url{https://github.com/mostafa-ja/sigma-binary} to ensure reproducibility and facilitate further research.

\subsection{Paper outline}
The remainder of the paper is organized as follows:
Section~\ref{sec:relatedwork} reviews related studies on evasion attacks and defenses. Section~\ref{sec:background} provides a review of the preliminaries. Section~\ref{sec:methods} details the proposed methods, including Prioritized Binary Rounding and $\sigma$-binary. Section~\ref{sec:experiments} describes the experimental setup and presents results. Section~\ref{sec:discussion} discusses the main findings, interprets the observed trade-offs, and highlights the limitations of the study. 
Section~\ref{sec:validity} analyzes the internal and external threats to validity and explains the measures taken to mitigate them. Finally, Section~\ref{sec:conclusion} presents the key findings and outlines directions for future research.

\section{Related Work}
\label{sec:relatedwork}

This section reviews two critical aspects of adversarial research in malware detection: the development of evasion attack strategies and the design of corresponding defense mechanisms. To provide a comprehensive overview, the discussion is divided into \textit{Evasion Attacks in Malware Detection} and \textit{Defenses Against Evasion Attacks}.

\subsection{Evasion Attacks in Malware Detection}

Adversarial attacks in Android malware detection can be broadly categorized into \textit{problem-space attacks}, which directly modify applications, and \textit{feature-space attacks}, which manipulate extracted feature representations. Both strategies exploit vulnerabilities in machine learning-based malware detectors while preserving the original functionality of the malware.

\textit{Problem-space attacks} alter application components such as the manifest file, Dalvik bytecode, or inject benign artifacts to bypass detection. Android HIV~\cite{chen2019android} perturbs Dalvik bytecode to evade detectors while maintaining functionality. Pierazzi et al.~\cite{pierazzi2020intriguing} introduced the injection of benign bytecode "gadgets" to fool detection models. Tang et al.~\cite{tang2024dapadv} expanded this with DapAdv, leveraging a hierarchical attention mechanism to optimize code slicing for adversarial repackaging. Advanced techniques include reinforcement learning-based attacks like HRAT~\cite{zhao2021structural}, and hybrid approaches using generative adversarial networks (GANs)~\cite{goodfellow2020generative} with coevolution algorithms~\cite{li2023black} have further improved the effectiveness of problem-space attacks. Black-box methods such as EvadeDroid~\cite{bostani2024evadedroid} demonstrate success with minimal feature knowledge, relying on random perturbations in benign code.

\textit{Feature-space attacks} focus on manipulating extracted feature vectors without altering the application itself. \mbox{Gradient-based} methods~\cite{huang2018adversarial, grosse2017adversarial} optimize perturbations to evade detection. Ensemble-based techniques such as those in ~\cite{li2020adversarial} combine diverse manipulation strategies for enhanced efficacy. GAN-based approaches, such as E-MalGAN~\cite{li2019adversarial} and the method proposed by Shahpasand et al.~\cite{shahpasand2019adversarial}, craft adversarial examples that evade both malware detectors and adversarial detection systems.

\subsection{Defenses Against Evasion Attacks}

Developing effective defenses against adversarial attacks has been a critical area of research, focusing on enhancing model robustness.

\textit{Adversarial training} is a well-established defense strategy that enhances model resilience by incorporating adversarial examples into the training process. While it proves effective against the perturbations encountered during training~\cite{grosse2017adversarial}, this approach faces limitations, including vulnerability to novel attack styles, high computational overhead, and a tendency to overfit to specific attack methods~\cite{carlini2019evaluating}.

\textit{Ensemble methods} aim to enhance robustness by combining multiple classifiers or detectors. Smutz and Stavrou~\cite{smutz2016tree} proposed mutual agreement analysis, flagging uncertain predictions based on classifier disagreement. Ficco~\cite{ficco2021malware} introduced a dynamic ensemble system combining generic and specialized detectors. However, applying ensemble strategies to deep learning models remains challenging due to difficulties in coordinating diverse model predictions.

Defenses with \textit{auxiliary models} utilize separate frameworks to identify adversarial examples or to purify inputs. Li et al.~\cite{li2021robust} employed a Variational Autoencoder (VAE) to detect adversarial malware via reconstruction errors, while Li et al.~\cite{li2018hashtran} combined hash transformations with denoising autoencoders to mitigate perturbations. Despite initial success, these approaches often struggle against adaptive attacks.

Recently, \textit{hybrid approaches} have gained prominence, combining multiple defense strategies to address the limitations of individual methods. For instance PAD-SMA~\cite{li2023pad} integrates a malware detection system with an adversary detector, both strengthened through adversarial training with a diverse range of attacks. While this approach enhances robustness and generalization, it also incurs significant computational overhead and exhibits lower accuracy on benign samples, limiting its practicality in real-world scenarios.

The ongoing advancements in both attack and defense mechanisms underscore the dynamic nature of this field, necessitating continuous innovation to address emerging threats effectively.

\section{Preliminaries}
\label{sec:background}

In this section, we introduce the notations used throughout the paper and describe the integrated malware and adversary detectors, focusing on the key concepts that form the basis of our work.

\subsection{Notations}

To formalize the concepts and methods discussed in this paper, we define the following key notations:

\begin{itemize}
    \item Let $Z$ represent the problem space, where each $z \in Z$ corresponds to an Android application sample.
    \item A feature extraction function $\phi: Z \to X$ maps samples from the problem space to a $d$-dimensional discrete feature-space $X \subset \mathbb{R}^d$.
    \item A malware detector $f: Z \to Y$ assigns a label $y \in Y = \{0, 1\}$, where $0$ represents benign and $1$ indicates malicious. The detector is defined as \( f(z) = \varphi_\theta(\phi(z)) \), where \( \varphi_\theta : X \to Y \) is a ML model with parameters $\theta$.
    \item An adversary detector $g: Z \to \mathbb{R}$ identifies adversarial examples. It operates in the feature-space as \mbox{$g(z) = \psi_\vartheta(\phi(z))$}, where $\psi_\vartheta$ is a secondary model with learnable parameters $\vartheta$. A sample is flagged as adversarial if $g(z) > \tau$, where $\tau$ is a predefined threshold.
    \item For a sample-label pair $(z, y)$, the feature representation is $x = \phi(z)$. An adversarial example is defined as $z' = z + \delta_z$, where $\delta_z$ denotes perturbations in the problem space. Correspondingly, $x' = \phi(z')$ represents the perturbed feature-space sample, expressed as $x' = x + \delta_x$ where $\delta_x$ denotes feature-space perturbations.
\end{itemize}

The loss function for the malware detector is denoted as \( \mathcal{L}_\varphi(\theta, x + \delta_x, y) \), capturing the discrepancy between the model's predictions and true labels.

\subsection{Integrated Malware and Adversary Detectors}

To enhance the robustness of malware detection, a secondary adversary detector $g$ is integrated with the primary malware detector $f$. The models are defined as.
\begin{align}
    f(z) &= \varphi _\theta(\phi(z)), \\
    g(z) &= \psi_\vartheta(\phi(z)),
\end{align}
where \( \theta \) and \( \vartheta \) are the parameter sets of the respective models. The prediction process is as follows:

\small
\begin{equation}
    \text{predict}(z) =
    \begin{cases} 
      f(z), & \text{if } g(z) \leq \tau, \\
      1, & \text{if } g(z) > \tau \text{ and } f(z) = 1, \\
      \text{further action}, & \text{if } g(z) > \tau \text{ and } f(z) = 0.
    \end{cases}
\end{equation}
\normalsize

When $g(z) > \tau$ and $f(z) = 0$, the sample is classified as suspicious and handled in one of two ways: (i) \textbf{Deferred}, where the input is excluded from classification and reserved for further analysis; or (ii) \textbf{Conservative}, where the input is labeled as malicious to prioritize safety. This integrated approach ensures robust detection while addressing uncertainties introduced by adversarial examples.

\subsection{Evasion Attacks}
Evasion attacks target both the problem and feature spaces. In the problem space, an adversarial sample $z'$ satisfies:
\begin{equation}
    f(z') = 0, \quad g(z') \leq \tau.
\end{equation}
In the feature-space, this translates to perturbations $x' = x + \delta_x$ such that:
\begin{equation}
    \varphi _\theta(x') = 0, \quad \psi_\vartheta(x') \leq \tau, \quad x' \in [\underline{u}, \overline{u}],
\end{equation}
The interval \( [\underline{u}, \overline{u}] \) defines the feature-space boundaries, where \( \underline{u}\) and \( \overline{u} \) represent the lower and upper limits, respectively. These constraints ensure that perturbations remain within the valid feature range, maintaining the feasibility of adversarial examples.

To address the disconnect between the feature-space and the problem space, an approximate inverse mapping function, \(\tilde{\phi}^{-1}\), is utilized \cite{vsrndic2014practical}. This function maps perturbations from the feature-space back to the problem space, ensuring practical feasibility.

\subsection{Adversarial Attack Methods}
\label{subsec:attack-methods}

We consider a comprehensive set of attacks spanning greedy, gradient-based, and ensemble strategies.  
These methods provide a diverse benchmark for assessing defense robustness in binary malware feature spaces.

\paragraph{Projected Gradient Descent (PGD).}  
PGD~\cite{madry2017towards} performs iterative gradient ascent under a constraint projection:
\[
\delta^{(t+1)} = \Pi_{\mathcal{C}}\big(\delta^{(t)} + \alpha \nabla_\delta F(\theta, x+\delta^{(t)}, y)\big),
\]
where \(\Pi_{\mathcal{C}}\) ensures feasibility, and the gradient is normalized under \(\ell_1\), \(\ell_2\), or \(\ell_\infty\) norms (PGD-\(\ell_p\)).  
PGD is widely adopted for its strong, iterative optimization performance.

\paragraph{Fast Gradient Sign Method (FGSM).}  
FGSM~\cite{goodfellow2014explaining} performs a single-step update along the sign of the gradient:
\[
x' = \operatorname{Proj}_{[\underline{u},\, \overline{u}]}\big(x + \varepsilon \cdot \operatorname{sign}(\nabla_x F(\theta, x, y))\big),
\]
where \(\varepsilon\) controls perturbation magnitude.  
It serves as a fast, baseline adversarial method and as a precursor to iterative variants.

\paragraph{Randomized FGSM (rFGSM).}  
rFGSM~\cite{huang2018adversarial} extends FGSM by introducing a random initialization and applying randomized rounding for discrete inputs, increasing attack diversity and success in binary domains.

\paragraph{Bit Gradient Ascent (BGA) and Bit Coordinate Ascent (BCA).}  
Both BGA and BCA~\cite{grosse2017adversarial,huang2018adversarial} target binary features directly.  
BGA flips all bits with gradients exceeding a dynamic threshold, while BCA greedily flips the single bit with the largest positive gradient at each step.  
These methods are computationally efficient yet highly effective for discrete domains.

\paragraph{Carlini--Wagner (CW) Attack.}  
The CW attack~\cite{carlini2017towards} minimizes the \(\ell_2\)-norm of perturbations while enforcing misclassification through constrained optimization, producing minimal yet effective feature modifications.

\paragraph{Mixture and Ensemble Attacks.}  
Mixture-based attacks~\cite{li2020adversarial,li2023pad} combine multiple attack strategies to enhance strength and transferability.  
The Max Mixture of Attacks (MaxMA) selects the most adversarial example among multiple candidates; Iterative MaxMA (iMaxMA) repeats this process iteratively; and the Stepwise Mixture of Attacks (SMA) applies a staged combination of attacks with adaptive step sizes and norms.

\paragraph{Mimicry Attack.}  
Mimicry attacks~\cite{biggio2013evasion,vsrndic2014practical} are gradient-free strategies that perturb malware to resemble benign samples.  
By iteratively querying the target model and adjusting features, they achieve evasion without requiring access to model internals.

\paragraph{\(\sigma\)-zero Attack} 
The \(\sigma\)-zero attack~\cite{cina2024sigma} is designed to minimize the number of modified features by approximating the \(\ell_0\)-norm and applying adaptive projections. It is effective in generating highly sparse perturbations, challenging defenses that assume adversarial examples require extensive feature modifications.

\subsection{Oblivious vs. Adaptive Attacks}

Adversarial attacks targeting malware detection systems can be broadly classified into oblivious and adaptive attacks, based on the attacker’s knowledge of the adversary detector~\( g \).

Oblivious attacks operate without the knowledge of the adversary detector \( g \). These attacks aim solely to bypass the malware detector \( f \) by ensuring that adversarial inputs are misclassified as benign. As such, oblivious attacks overlook the presence of \( g \), making them less effective when \( g \) is actively present. The adversary detector \( g \) can flag such inputs as adversarial, limiting their ability to evade the overall detection framework.

Adaptive attacks, on the other hand, explicitly account for the adversary detector \( g \), represented as \( g(z') = \psi_\vartheta(\phi(z')) \). These attacks require that adversarial input not only evades the malware detector \( f \) but also satisfies \( g(z') \leq \tau \). This dual requirement makes adaptive attacks more complex and challenging to defend against.

The optimization problem for adaptive attacks is formulated~as follows:
\small
\begin{equation}
\min_{x' \in [\underline{u}, \overline{u}]} \mathcal{L}_\varphi(\theta, x', 0) \quad \text{s.t.} \quad \psi_\vartheta(x') \leq \tau \quad \text{and} \quad x' \in X.
\end{equation}
\normalsize

This formulation substitutes the condition \(  \varphi_\theta(x') = 0 \) with the minimization of \( \mathcal{L}_\varphi(\theta, x', 0) \), addressing the issue of non-differentiability.

A critical challenge in this optimization process lies in the nonlinear nature of the constraint $\psi_\vartheta(x') \leq \tau$, which precludes the use of standard gradient descent techniques. Unlike oblivious attacks, the problem cannot be reformulated by simply swapping the objective and constraint. The non-linear constraint imposed by $g$ requires a more sophisticated approach to ensure that adversarial inputs evade both $f$ and $g$ while adhering to distortion bounds.  

To address this difficulty, existing literature adopts a Lagrangian relaxation approach, previously applied to constructing minimum-distortion adversarial examples \cite{carlini2017towards}. The optimization objective is reformulated as follows.  
\begin{equation}
    \min_{x' \in [\underline{u}, \overline{u}]} \mathcal{L}_\varphi(\theta, x', 0) 
     + \lambda \psi_\vartheta(x')
\label{eq:my_equation}
\end{equation}
where $\lambda \geq 0$ is a penalty factor that adjusts the relative importance of evading $f$ versus reducing the response of $g$. The optimal value of $\lambda$ is determined through a binary search, ensuring an effective trade-off between these competing objectives \cite{bryniarski2021evading}.

This penalty-based formulation enables adaptive attacks to effectively bypass both $f$ and $g$, while addressing the inherent complexities of optimizing over a constrained, non-linear adversarial detection model.

\subsection{Binary Rounding Methods}

Binary rounding is essential for generating adversarial examples that respect the binary feature constraints inherent in malware detection. During optimization, gradient-based methods often yield intermediate solutions with continuous feature values, violating these constraints. To address this, previous studies have explored two primary approaches:

\begin{itemize}
    \item \textbf{Deterministic Binary Rounding:} This method applies a fixed threshold—typically 0.5—to convert continuous values into binary features. Values equal to or greater than 0.5 are rounded to 1, while values below 0.5 are rounded to 0.
    \item \textbf{Randomized Binary Rounding:} This probabilistic approach rounds continuous values based on their magnitude. For example, a value of 0.7 is rounded to 1 with a probability of 70\% and to 0 with a probability of 30\%.
\end{itemize}

\section{Proposed Approach}
\label{sec:methods}

This section presents the proposed methods for crafting adversarial examples in binary feature spaces. The approach addresses two key challenges: (i) ensuring perturbations conform to binary constraints and (ii) achieving dual objectives—evading malware detection and bypassing adversarial defense mechanisms.

\subsection{Threat Model and Design Objective}
\label{subsection:Threat_Model}
We assume a white-box attack setting, where the attacker has full knowledge of both the malware detector $f$ and the adversary detector $g$, including their architectures, trained parameters, and feature representations \cite{carlini2017towards, biggio2018wild}. This assumption allows for worst-case evaluations of the robustness of ML-based defenses against adversarial attacks.

The attacker’s objective is to generate adversarial examples that evade $f$ while remaining undetected by $g$. An approximate inverse mapping function $\tilde{\phi}^{-1}$ is employed, as proposed in prior studies \cite{li2023pad, li2020adversarial}, to translate feature-space perturbations into the problem space, ensuring the adversarial examples retain their functional validity.

It is worth noting that in oblivious attacks, the attacker is unaware of the existence of \( g \) and, consequently, does not consider \( g \) in the process of generating adversarial examples.

\subsection{Incorporating Distance Metrics}

An appropriate distance metric is essential for quantifying the similarity between original and adversarial samples. In binary feature spaces, the \textit{Hamming distance($d_H$)} is a natural choice, capturing the number of differing features between two binary vectors:
\begin{align}
    d_H(x, x') = \sum_{i=1}^d \mathbb{I}(x_i \neq x'_i).
\end{align}
Where $\mathbb{I}(\cdot)$ is the indicator function. The Hamming distance aligns directly with the $\ell_0$-norm for binary vectors and serves as a structured and interpretable measure of dissimilarity.

\subsection{Problem Formulation}
The adversarial attack is formulated as an optimization problem in the binary feature space. For a given malware instance-label pair $(x, y)$, where $x = \phi(z)$ and $y = 1$, the attacker seeks to determine the minimal feature modification $\delta^\star$, quantified by the Hamming distance $d_H$, such that the resulting adversarial example $x^\star = x + \delta^\star$ satisfies.
\begin{align}
    \delta^\star \in \arg & \min_\delta d_H(x, x + \delta), \\
    \text{s.t.} & \quad \varphi_\theta(x + \delta) = 0, \label{eq:phi_constraint} \\
    & \quad \psi_\vartheta(x + \delta) \leq \tau, \label{eq:psi_constraint} \\
    & \quad x + \delta \in \mathcal{X}, \label{eq:space_constraint} \\
    & \quad x + \delta \in [\underline{u}, \overline{u}]. \label{eq:box_constraint}
\end{align}
Due to the intractability of solving this problem directly, we reformulate it using a smooth surrogate objective:
\begin{align}
    \delta^\star \in \arg & \min_\delta \mathcal{L}(\theta, \vartheta, x + \delta) + \frac{1}{d} \widetilde {d}_H(x, x + \delta),  \\
    &\quad \text{s.t.} \quad x + \delta \in \mathcal{X}, \quad x + \delta \in [\underline{u}, \overline{u}],
\end{align}
Where \( \widetilde{d}_H(x, x + \delta) \) represents a differentiable approximation of the Hamming distance, normalized by the number of features \( d \) to ensure its value lies within the interval \([0, 1]\). This normalization removes the need for hyperparameter tuning to balance the trade-off between the loss and the perturbation size, thereby avoiding computationally expensive line searches for each input sample \cite{cina2024sigma}.

The total loss \( \mathcal{L} \) is defined as follows.
\small
\begin{align}
    \mathcal{L}(\theta, \vartheta, x + \delta) 
    &= \mathcal{L}_\varphi(\theta, x + \delta, 0) + C \cdot \mathcal{L}_\psi(\vartheta, x + \delta).
\end{align}
\normalsize
where:
\small
\begin{align}
    &\mathcal{L}_\varphi(\theta, x + \delta, 0) = 
    \max(\varphi_\theta^1(x + \delta) - \varphi_\theta^0(x + \delta), -\kappa_1).
\end{align}
\begin{align}
    \mathcal{L}_\psi(\vartheta, x + \delta) = 
    \max(\psi_\vartheta(x + \delta) - \tau, -\kappa_2).
\end{align}
\normalsize

\noindent In these equations:
\begin{itemize}
    \item \( \mathcal{L}_\varphi(\theta, x + \delta, 0) \) represents the loss function of the malware detector \( f \), with a target label of zero.
    \item \( \mathcal{L}_\psi(\vartheta, x + \delta) \) represents the loss function of the adversary detector  \( g \).
    \item \( \varphi_\theta^k \) represents the logit output of \( f \) for class \( k \in \{0, 1\} \).
    \item \( C \geq 0 \)  is a penalty weight optimized via binary search to balance the loss contributions of \( f \) and \( g \).
    \item \( \kappa_1 \) and \( \kappa_2 \) are confidence margins for \( f \) and \( g \), ensuring robust misclassification and avoidance of detection.
\end{itemize}

We define the loss for the malware detector \( f \) using the differences between logits, a method whose rationale is extensively discussed in \cite{carlini2017towards}. Furthermore, independently bounding the loss functions ensures that minimizing one loss term does not inherently satisfy the constraints for both detectors, thereby preserving the distinct objectives of \( f \) and \( g \).

\subsubsection{Hamming Distance Approximation}
To enable gradient-based optimization, the Hamming distance $d_H$ is approximated as follows.
\begin{align}
    \widetilde{d}_H(x, x + \delta) = \sum_{i=1}^d \frac{\delta_i^2}{\delta_i^2 + \sigma}.
\label{eq:Distance_Approximation}
\end{align}

Here, $\sigma > 0$ serves as a smoothing parameter to ensure differentiability. However, this approximation may lead to non-sparse solutions. To mitigate this, an adaptive projection operator \( \Pi_{\gamma} \) enforces sparsity by setting perturbation components below a threshold \( \gamma \) to zero, promoting both sparsity and adversarial effectiveness \cite{cina2024sigma}.

\begin{algorithm}[tbp]

\caption{$\sigma$-binary Attack with Prioritized Binary Rounding}
\label{alg:sigma_binary_attack}
\KwIn{
\small
$x \in [0, 1]^d$: input sample; \\
$\theta$: malware detector \(f\); $\vartheta$: adversary detector \(g\); \\
$N, N_{\text{binary\_search}}$: max iterations; $\eta_0$: initial step size; \\
$\sigma$: smoothing parameter; $C_{\text{init}}$: initial penalty factor; \\
$\gamma_0$: initial sparsity threshold; $t$: sparsity adjustment factor; \\
$\kappa_{\text{1,init}}, \kappa_{\text{2,init}}$: initial confidence values; \\
$\alpha_1, \alpha_2$: confidence bounds; $\epsilon$: loss convergence threshold.
}
\normalsize
\KwOut{$x^\star$: binary adversarial example.}
\small 
\texttt{Loss}$(\theta, \vartheta, x, \delta) = \mathcal{L}(\theta, \vartheta, x + \delta) + \frac{1}{d} \widetilde{d}_H(x, x + \delta)$\;
\normalsize
Initialize: $\delta^\star \gets \infty$, $C \gets C_{\text{init}}$, \texttt{\small best\_dist} $\gets \infty$;
\For{$\text{OuterStep} \gets 1$ \textbf{to} $N_{\text{binary\_search}}$}{
    Initialize: $\delta \gets 0$, $\gamma \gets \gamma_0$, $\eta \gets \eta_0$, 
    $\kappa_1 \gets \kappa_{\text{1,init}}$, $\kappa_2 \gets \kappa_{\text{2,init}}$\; 
    \For{$i \gets 1$ \textbf{to} $N$}{
        $\nabla g \gets \nabla_\delta \texttt{\small Loss}(\theta, \vartheta, x, \delta)$\;
        
        $\nabla g \gets \frac{\nabla g}{\|\nabla g\|_\infty}$\;

        $\eta \gets \text{cosine\_annealing}(\eta_0, i)$\;
        $\delta \gets \text{clip}(x + (\delta - \eta \cdot \nabla g)) - x$\;
        $\delta \gets \Pi_\gamma(\delta)$\;
        \If{$(x+\delta)$ evades both \( f \) \& \( g \)}{
            $\gamma \gets \gamma + t \cdot \eta$\;
        } \Else{
            $\gamma \gets \gamma - t \cdot \eta$\;
        }
        $\delta_\text{binary} \gets R_\text{prioritized\_binary}(x, x+\delta, \theta, \vartheta)$\;
        \If{$d_H(\delta_\text{binary}, 0) < \texttt{\small best\_dist}$ \textbf{and}\\ $(x+\delta_\text{binary})$ evades both \( f \) \& \( g \)}{
            $\delta^\star \gets \delta_\text{binary}$\;
            \texttt{\small best\_dist} $\gets d_H(\delta_\text{binary}, 0)$\;
        }
        \If{$|\texttt{\small Loss} - \texttt{\small previous\_Loss}| < \epsilon$}{
            \If{$(x+\delta)$ evades both \( f \) \& \( g \)}{
                $\kappa_1 \gets -\alpha_1$, $\kappa_2 \gets -\alpha_2$\;
            } \Else{
                $\kappa_1 \gets \alpha_1$, $\kappa_2 \gets \alpha_2$\;
            }
        }
    }
    Update $C$ via binary search\;
}
\textbf{Return:} $x^\star \gets x + \delta^\star$\;

\end{algorithm}

\normalsize

\subsubsection{Binary Rounding}
To convert continuous solutions into binary values, we employ \textit{Prioritized Binary Rounding} (detailed in Subsection~\ref{subsection:Prioritized Binary Rounding}), which prioritizes feature modifications based on their impact on model decisions. This method ensures compliance with binary constraints while minimizing perturbations.

\subsubsection{Dynamic Confidence Adjustment}
In adversarial optimization, the solution typically converges to an optimal result in the continuous-valued space. However, during the binary rounding process, where continuous values are discretized into binary form, suboptimality can be introduced. This suboptimality arises because the process of forcing discrete values can disrupt the fine-tuned structure of the continuous solution.
To address this issue, we employ the \textit{Dynamic Confidence Adjustment} method. After the initial convergence of the optimization process, the confidence parameter \(\kappa\) is dynamically adjusted within a bounded range \([-\alpha, \alpha]\). This adjustment enables the optimization process to explore regions near the decision boundary, facilitating the discovery of a binary solution that more effectively satisfies the adversarial objectives.

\subsubsection{Algorithm Details for the \texorpdfstring{\(\sigma\)}{sigma}-binary Attack}

The $\sigma$-Binary Attack, outlined in Algorithm~\ref{alg:sigma_binary_attack}, generates binary adversarial examples with minimal perturbations to evade the malware detector \(f\) and, if applicable, the adversary detector \(g\). Key parameters, such as the penalty factor \(C\), sparsity threshold, and confidence margins \(\kappa_1\) and \(\kappa_2\), are initialized with \(\kappa_1\) and \(\kappa_2\) set to small values to prioritize early boundary crossing for effective evasion.

The algorithm employs two nested loops: an outer binary search loop to adjust \(C\) for balancing loss terms, and an inner optimization loop to minimize a loss function combining the evasion objective and a smooth approximation of the Hamming distance. Perturbations are iteratively updated via a cosine annealing learning rate, clipped gradient descent, and enforced sparsity. Non-binary perturbations are converted to binary form using \textit{PBR} to maintain functionality and input constraints.

Once the total loss converges, the thresholds are dynamically adjusted throughout the optimization process based on evasion success. The algorithm ultimately outputs the best adversarial example, denoted \( x^\star = x + \delta^\star \), which minimizes the Hamming distance while successfully bypassing detection.

\subsection{Prioritized Binary Rounding}
\label{subsection:Prioritized Binary Rounding}

To efficiently map continuous perturbations to binary feature vectors, the proposed \(\sigma\)-binary attack framework introduces a novel technique termed \textit{PBR}. This method enforces binary constraints while minimizing unnecessary feature modifications, achieving a balance between adversarial effectiveness and perturbation sparsity.

Conventional binary rounding approaches typically rely on uniform deterministic or randomized thresholds, thereby neglecting variations in feature importance and their respective impact on the model’s decision. In contrast, PBR selects features for modification according to a principled ordering that combines two criteria:
\begin{itemize}
    \item \textbf{Perturbation magnitude:} Features exhibiting larger deviations from their original values are given higher priority for modification.
    \item \textbf{Feature importance:} Features with greater influence on the model’s decision boundary are prioritized to strengthen the adversarial effect.
\end{itemize}

In the proposed implementation, features are primarily ranked by their gradient-derived perturbation magnitude. 
When multiple features exhibit similar magnitudes, \emph{discriminative feature importance}—computed on the training dataset—is used as a secondary tie-breaking criterion. 
This auxiliary ranking improves rounding stability by prioritizing features that are statistically more informative for classification.  
Discriminative importance is estimated from the frequency of feature occurrence across classes, where features appearing predominantly in either benign or malicious samples are considered more discriminative.  
Consequently, the PBR mechanism prioritizes features that exhibit strong gradient magnitudes and high discriminative relevance to classification outcomes.
The algorithm proceeds as follows.

\begin{algorithm}[tbp]
\caption{Prioritized Binary Rounding}
\label{alg:binary_rounding}
\KwIn{
$\mathbf{x}_{\text{orig}} \in [0, 1]^d$: original input; \\ 
$\mathbf{x}_{\text{adv}} \in [0, 1]^d$: adversarial input; \\ 
$f_\theta$: target model; $\tau$: perturbation threshold.
}
\KwOut{$\mathbf{x}^\star$: binary adversarial example.}

Initialize: $\mathbf{x}^\star \gets \mathbf{x}_{\text{orig}}$\;

\If{$f_\theta(\mathbf{x}^\star)$ is benign}{ 
    \Return $\mathbf{x}^\star$\; 
}

Compute perturbation magnitude: $\boldsymbol{\Delta} \gets |\mathbf{x}_{\text{adv}} - \mathbf{x}_{\text{orig}}|$\;

Generate mask for significant perturbations: $\mathbf{M} \gets \mathbbm{1}(\mathbf{\Delta} > \tau)$\;

Sort indices by descending perturbation magnitude: \\
\hspace{1.5em} $\text{FeatureOrder} \gets \text{argsort}(\mathbf{\Delta}, \text{desc})$\;

\For{$i \gets 1$ \textbf{to} $\sum(\mathbf{M})$}{
    Update feature $j \gets \text{FeatureOrder}[i]$: 
    \[
    \mathbf{x}^\star[j] \gets 
    \begin{cases} 
    1, & \text{if } \mathbf{x}_{\text{adv}}[j] > \mathbf{x}_{\text{orig}}[j], \\ 
    0, & \text{otherwise.}
    \end{cases}
    \]

    \If{$f_\theta(\mathbf{x}^\star)$ is benign}{ 
        \Return $\mathbf{x}^\star$\; 
    }
}

\Return $\mathbf{x}^\star$\;

\end{algorithm}

First, the perturbation magnitude for each feature is computed as the absolute difference between the adversarial and original feature values. Features with perturbations exceeding a predefined threshold $\tau$ are marked as candidates for modification using a binary mask. Candidate features are sorted in descending order based on perturbation magnitude and relative importance, determined by their position in the feature space, to prioritize the most impactful changes.

Each selected feature is iteratively updated by rounding it to one if the adversarial feature value is greater than the original; otherwise, it is set to zero. After each modification, the adversarial example is evaluated to determine whether it successfully evades detection by the malware detector. If the detector classifies the sample as benign, the rounding process terminates early, and the rounded result is returned. If early termination does not occur, the algorithm continues updating all significant features before outputting the final binary adversarial example.

The joint use of $\sigma$-binary and PBR enables both fine-grained optimization and faithful binary constraint enforcement.
This dual mechanism forms the foundation for the empirical evaluations presented in Section \ref{sec:experiments}.

\section{Experiments}
\label{sec:experiments}

We evaluate the proposed methods through an extensive experimental campaign designed to: (i) compare our adversarial attack against prior approaches, (ii) assess the robustness of multiple defense mechanisms under diverse threat models, and (iii) quantify trade-offs between robustness, standard classification performance, and computational cost. The evaluation is organized around five research questions (RQs):

\begin{itemize}
    \item \textbf{RQ1 — Performance of the \(\sigma\)-binary attack.} How does the \(\sigma\)-binary attack—enhanced through the PBR mechanism—perform compared to existing adversarial attack strategies in binary feature spaces?
    \item \textbf{RQ2 — Baseline performance of defenses.} What are the detection accuracies and false alarm characteristics of the evaluated defenses under standard (non-adversarial) conditions?
    \item \textbf{RQ3 — Robustness to oblivious attacks.} How resilient are defenses when the adversary is unaware of the adversary detector \(g\) (i.e., the oblivious threat model)?
    \item \textbf{RQ4 — Robustness to adaptive attacks.} How do defenses withstand adaptive attacks that explicitly account for both the malware detector \(f\) and the adversary detector \(g\), representing a worst-case threat model?
    \item \textbf{RQ5 — Trade-offs: robustness, accuracy, and computational cost.} What are the empirical trade-offs between adversarial robustness, standard (non-adversarial) classification accuracy, and the computational cost of each defense strategy?
\end{itemize}

\subsection{Experimental framework}
\label{subsec:experimental_design}

This subsection summarizes the dataset, feature extraction pipeline, defenses, and attacks evaluated, metrics, and experimental configuration used throughout the study.

\subsubsection{Dataset}
\label{subsubsec:Dataset}
We utilized the \textit{MalScan} dataset \cite{wu2019malscan}, a widely recognized collection comprising 30,715 Android applications, with 15,430 malicious and 15,285 benign samples. The dataset spans applications from 2011 to 2018 (Table~\ref{tab:datasets}). The near-balanced class distribution mitigates training bias, while the eight-year span of samples captures temporal variability in benign and malicious behaviors.

\begin{table}[ht]
\centering
\caption{Dataset summary (MalScan).}
\label{tab:datasets}
\small
\begin{tabular}{lccc}
\toprule
\textbf{Year} & \textbf{Benign} & \textbf{Malware} & \textbf{Total} \\
\midrule
2011 & 1,920 & 1,916 & 3,836 \\
2012 & 1,875 & 2,000 & 3,875 \\
2013 & 1,896 & 2,000 & 3,896 \\
2014 & 1,826 & 1,982 & 3,808 \\
2015 & 1,811 & 1,839 & 3,650 \\
2016 & 2,015 & 1,940 & 3,955 \\
2017 & 1,884 & 1,834 & 3,718 \\
2018 & 2,058 & 1,919 & 3,977 \\
\midrule
\textbf{Total} & \textbf{15,285} & \textbf{15,430} & \textbf{30,715} \\
\bottomrule
\end{tabular}
\end{table}

\subsubsection{Feature Extraction}
Feature extraction was performed using Drebin \cite{arp2014drebin}, which analyzes Android application packages (APKs) to construct a binary feature space. Features were extracted from the Android manifest and the disassembled DEX code using the Androguard tool. These features were organized into eight categories, such as hardware components, permissions, intents, API calls, and class names. Each APK is represented as a binary feature vector, where each dimension indicates the presence (\(1\)) or absence (\(0\)) of a specific feature. Consistent with prior work~\cite{li2023pad}, we exclude easily modifiable features (e.g., package names) and retain the 10,000 most frequent features, ensuring robust representation for evaluation.

\subsubsection{Defenses Considered for Comparative Analysis}
\label{subsubsection:defenses}

In this study, we evaluate a range of defenses drawn from prior research, categorized into three distinct types:  
\textit{(i)} methods that enhance the resilience of malware detectors through adversarial training,  
\textit{(ii)} methods that integrate malware detection with adversarial example detection mechanisms, and  
\textit{(iii)} methods that combine both adversarial training and adversary detection mechanisms for comprehensive defense.  

Below, we provide a detailed overview of each defense.  

\begin{itemize}
    \item \textbf{DNN}~\cite{grosse2017adversarial}: A baseline deep neural network model for malware detection. This model does not incorporate specific countermeasures against evasion attacks and serves as a reference for comparing the robustness of advanced defense mechanisms.

    \item \textbf{DNN\(^+\)}~\cite{grosse2017statistical}: An extension of the DNN model that incorporates a secondary detector. This detector introduces an additional outlier class to identify adversarial examples.

    \item \textbf{KDE}~\cite{pang2018towards}: Combines a DNN with a Kernel Density Estimation (KDE)-based secondary detector. The KDE mechanism identifies adversarial examples by analyzing deviations in activations of the DNN's penultimate layer.

    \item \textbf{DLA}~\cite{sperl2020dla}: Dense Layer Analysis (DLA) incorporates a secondary detector alongside the DNN. By evaluating activations across all dense layers, DLA distinguishes between normal and adversarial examples, adding a robust layer of defense.

    \item \textbf{AT-rFGSM\(^k\)}~\cite{huang2018adversarial}: A defense mechanism that strengthens the DNN through adversarial training using the FGSM\(^k\) attack. This approach employs randomized rounding projections to increase the model's resilience to adversarial perturbations.

    \item \textbf{AT-MaxMA}~\cite{li2020adversarial}: Enhances DNN robustness by incorporating adversarial training with a Mixture of Attacks (MaxMA) strategy. This method combines multiple attack techniques to construct a more comprehensive and resilient defense.

    \item \textbf{ICNN}~\cite{li2023pad}: Integrates a DNN with an Input Convex Neural Network (ICNN) as a secondary detector. The ICNN operates independently of the DNN's architecture and analyzes the feature-space to detect adversarial examples.

    \item \textbf{PAD-SMA}~\cite{li2023pad}:  The PAD framework (Principled Adversarial Malware Detection) integrates a DNN-based malware detector with an ICNN-based adversary detector. Both components are fortified through adversarial training using the Stepwise Mixture of Attacks (SMA), making this approach a representative example of defenses that combine adversarial training with adversary detection mechanisms.
\end{itemize}

\subsubsection{Evasion attack methods for comparative analysis}
\label{subsec:exp-attack-methods}

To assess the effectiveness of the proposed \(\sigma\)-binary attack, we compare it against a representative and diverse suite of established evasion techniques encompassing gradient-based, ensemble, gradient-free, and optimization-driven approaches. The selection follows established benchmarking practice~\cite{li2023pad,li2020adversarial} and augmented with sparsity-oriented solvers for a more comprehensive analysis. Details on implementation are provided in Section~\ref{subsec:attack-methods}. The comparative set comprises:

\begin{itemize}
  \item \textbf{Gradient-based methods:} discrete-tailored and continuous gradient approaches, including Bit Coordinate Ascent (BCA), Bit Gradient Ascent (BGA), the Grosse gradient-directed addition, randomized FGSM (rFGSM), and PGD variants (PGD-\(\ell_1\), PGD-\(\ell_2\), PGD-\(\ell_\infty\)). Continuous outputs are converted to binary vectors via deterministic or randomized rounding as appropriate.
  \item \textbf{Ensemble / mixture attacks:} Iterative MaxMA (iMaxMA) and Stepwise Mixture of Attacks (SMA), which combine multiple primitives to improve transferability~\cite{li2020adversarial,li2023pad}.
  \item \textbf{Gradient-free methods:} Mimicry-style attacks that modify malware features to resemble benign instances without requiring gradient access~\cite{vsrndic2014practical,biggio2013evasion}.
  \item \textbf{Optimization / sparsity-oriented:} Carlini–Wagner (CW, \(\ell_2\)) and the \(\sigma\)-zero solver targeting \(\ell_0\)-like sparsity~\cite{carlini2017towards,cina2024sigma}.
\end{itemize}

Including both CW and \(\sigma\)-zero permits a direct comparison between our binary-targeted \(\sigma\)-binary attack and established distance-optimized methods that explicitly minimize perturbation magnitude.

\subsubsection{Evaluation Metrics}

To rigorously assess the performance and robustness of defenses, we utilize a comprehensive suite of evaluation metrics. These metrics capture both the classification accuracy and resilience to adversarial attacks, and are defined as follows.

\begin{itemize}
    \item \textbf{False Positive Rate (FPR):} The proportion of benign examples that are incorrectly classified as malicious. This reflects the model's tendency to overestimate threats.
    \item \textbf{False Negative Rate (FNR):} The proportion of malicious examples that are mistakenly classified as benign, indicating the model's vulnerability to evasion by malware.
    \item \textbf{Accuracy (Acc):} The overall ratio of correctly classified examples to the total number of samples, representing the model's general predictive capability.
    \item \textbf{F1 Score:} The harmonic mean of precision and recall, offering a robust evaluation for imbalanced datasets by balancing the trade-off between false positives and false negatives.
    \item \textbf{Attack Success Rate (\(\text{ASR}_k\)):} The proportion of adversarial attacks that successfully evade detection, where the Hamming distance between the original and adversarial samples satisfies \( d_H(\mathbf{x}, \mathbf{x}^\star) \leq k \). This metric measures the model's resistance to perturbations of varying magnitudes.
    \item \textbf{Median Perturbation (\(\tilde{d}_{H, \text{median}}\)):} The median Hamming distance of successful adversarial examples, quantifying the typical modification magnitude required to deceive the detector.
    \item \textbf{Interquartile Range (IQR)}: The dispersion of Hamming distances among successful attacks, reflecting the stability and consistency of perturbation magnitudes.
    \item \textbf{Excluded Samples (ExS):} For deferred detection models, this metric represents the fraction of inputs that are deferred by the adversary detector \( g \).
\end{itemize}

These metrics collectively provide a robust framework for evaluating both the predictive accuracy and adversarial robustness of the proposed defenses.

\subsection{Experimental setup}
\label{subsection:experimental_setup}

\subsubsection{Hardware, software and reproducibility}
Experiments were executed on a workstation running Microsoft Windows 11 Pro with an 11th Gen Intel Core i5-11500H CPU (6 cores, 12 threads), 32\,GB RAM, and an NVIDIA RTX A2000 Laptop GPU (4\,GB VRAM). The experimental codebase uses Python 3.12.5 and PyTorch 2.5.0 with CUDA 12.4. To facilitate reproducibility, all software dependencies, hyperparameters, and processed datasets (including the exact train/validation/test splits) are publicly available.

\subsubsection{Data splits and protocol}
All experiments use the MalScan dataset with fixed partitions: 60\% training, 20\% validation, and 20\% test. Identical splits and model initializations were maintained across experiments to ensure fair comparisons. Performance is reported on the full test set, which remained strictly held-out during training and validation.

\subsubsection{Model selection and hyperparameters}
Where available, defense models adopt the hyperparameter settings reported in their original works. This choice avoids confounding differences due to re-tuning and enables a direct assessment of comparative robustness under the conditions validated by prior studies.

\subsubsection{Numerical stability and bias control}
To limit numerical instability, gradient clipping was applied during training. Gradient-based attacks use normalized step sizes to ensure consistent perturbation scaling across defenses. Experimental bias was further reduced by fixing random seeds for Python, NumPy, and PyTorch.

\subsubsection{Attack configuration}
To evaluate the effectiveness of gradient-based attacks, we generate adversarial examples using all 3,104 malicious samples in the test dataset. For attack-strength comparisons, we allowed up to 1{,}000 gradient-descent iterations; for robustness evaluation we extended the budget to 10{,}000 iterations of gradient descent, ensuring a thorough exploration of the attack surface. While convergence often occurs earlier, this approach allows for a comprehensive evaluation~\cite{carlini2017towards}. For the CW attack, 8 iterations of binary search are conducted to optimize the penalty weight c, maximizing the impact of adversarial perturbations. Continuous outputs from gradient-based attacks were converted to the binary domain via deterministic rounding (threshold \(0.5\)), except where attacker formulations natively operate over binary variables. For the \(\sigma\)-binary attack we apply the PBR scheme to preserve adversarial efficacy while enforcing binary constraints.

\subsubsection{Detector calibration}
For defenses equipped with adversary detectors, the process is adjusted to use 4 iterations of binary search to identify the smallest effective penalty factor \(c\). Additionally,  thresholds for identifying suspicious samples are computed by excluding the top 5\% of validation examples with the highest confidence, as recommended by \cite{li2023pad} and \cite{sperl2020dla}. This strategy ensures that the thresholds are both practical and minimally disruptive to benign classification performance.

The above configuration establishes a standardized and reproducible evaluation pipeline for all experiments. The following section presents empirical results addressing RQ1–RQ5, highlighting the comparative robustness and efficiency of the proposed $\sigma$-binary framework.

\subsection{RQ1: Performance of \texorpdfstring{\(\sigma\)}{sigma}-binary attack}
\label{subsection:sigma-binary attack}

\definecolor{rowgray}{gray}{0.95}

\begin{table*}[t]
    \centering
    \caption{Attack Success Rate (ASR, \%) under four defense models. 
    Columns report ASR for 10, 20, 50, 100, and $\infty$ perturbation budgets, along with 
    mean runtime per sample $\bar{t}$ (in seconds/sample) and maximum VRAM usage (in GB). 
    The best result per row is in \textbf{bold}.}
    \vspace{2mm}
    \setlength{\tabcolsep}{4pt}
    \renewcommand{\arraystretch}{1.1}
    \small

    \begin{tabular}{l|ccccccc|ccccccc}
        \toprule
        \multirow{2}{*}{\textbf{Attack}} &
        \multicolumn{7}{c|}{\textbf{DNN}} &
        \multicolumn{7}{c}{\textbf{AT-rFGSM\(^k\)}} \\
        \cmidrule(lr){2-15}
         & ASR$_{10}$ & ASR$_{20}$ & ASR$_{50}$ & ASR$_{100}$ & ASR$_{\infty}$ & $\bar{t}$ & VRAM
         & ASR$_{10}$ & ASR$_{20}$ & ASR$_{50}$ & ASR$_{100}$ & ASR$_{\infty}$ & $\bar{t}$ & VRAM \\
        \midrule
        
        \rowcolor{rowgray} BCA & 93.2 & 99.7 & 99.8 & 99.8 & 99.8 & 0.008 & \textbf{0.6} &
        29.5 & 48.9 & 67.4 & 73.3 & 75.2 & \textbf{0.011} & \textbf{0.9} \\
        
        BGA & 2.1 & 2.1 & 2.1 & 2.1 & \textbf{100.0} & \textbf{0.001} & 0.8 &
        1.6 & 4.8 & 16.9 & 37.5 & 50.0 & 0.015 & 1.0 \\
        
        \rowcolor{rowgray} Grosse & 93.1 & 99.7 & 99.8 & 99.8 & 99.8 & 0.009 & \textbf{0.6} &
        29.5 & 49.1 & 67.5 & 73.7 & 75.6 & 0.012 & \textbf{0.9} \\
        
        rFGSM & 2.1 & 2.1 & 2.1 & 26.0 & 99.7 & 0.008 & 0.7 &
        1.1 & 1.2 & 3.4 & 14.3 & 79.5 & 0.011 & \textbf{0.9} \\
        
        \rowcolor{rowgray} PGD-$\ell_1$ & 93.2 & 99.7 & 99.8 & 99.8 & 99.8 & 0.012 & 0.7 & 
        39.1 & 57.3 & 73.8 & 78.3 & 79.5 & 0.018 & \textbf{0.9} \\
        
        PGD-$\ell_2$ & 2.1 & 2.1 & 2.1 & 2.1 & 93.8 & 0.012 & 0.7 &
        1.1 & 1.2 & 4.1 & 21.2 & 77.7 & 0.012 & \textbf{0.9} \\
        
        \rowcolor{rowgray} PGD-$\ell_\infty$ & 2.1 & 2.1 & 2.1 & 2.1 & 99.8 & 0.012 & 0.7 &
        1.1 & 1.2 & 4.4 & 20.6 & 84.5 & 0.012 & \textbf{0.9} \\
        
        iMaxMA & 2.1 & 2.1 & 2.1 & 2.1 & 99.8 & 0.036 & 0.8 &
        1.1 & 1.1 & 4.0 & 20.1 & 88.8 & 0.045 & 1.0 \\
        
        \rowcolor{rowgray} SMA & 93.2 & 99.7 & 99.8 & 99.8 & 99.8 & 0.011 & 0.8 &
        39.2 & 57.7 & 70.9 & 72.3 & 72.4 & 0.025 & 1.1 \\

        Mimicry & 16.6 & 35.1 & 77.8 & 98.4 & 99.8 & 0.012 & 1.0 &
        12.2 & 23.1 & 47.3 & 62.3 & 73.6 & 0.011 & 1.2 \\
        
        \rowcolor{rowgray} $\sigma$-zero & 89.4 & 94.7 & 94.9 & 94.9 & 94.9 & 0.028 & 0.8 &
        42.8 & 56.4 & 70.0 & 73.7 & 73.8 & 0.027 & \textbf{0.9} \\
        
        CW & 43.7 & 51.4 & 51.6 & 51.6 & 51.6 & 0.107 & 1.9 &
        20.8 & 29.9 & 38.7 & 43.2 & 44.6 & 0.105 & 2.2 \\
        
        \rowcolor{rowgray} $\sigma$-binary & \textbf{94.5} & \textbf{99.8} & \textbf{100.0} & \textbf{100.0} & \textbf{100.0} & 0.046 & 1.0 &
        \textbf{48.2} & \textbf{67.0} & \textbf{90.3} & \textbf{98.8} & \textbf{99.4} & 0.067 & 1.2 \\
        \bottomrule
    \end{tabular}

    \vspace{3mm} 

    \begin{tabular}{l|ccccccc|ccccccc}
        \toprule
        \multirow{2}{*}{\textbf{Attack}} &
        \multicolumn{7}{c|}{\textbf{AT-MaxMA}} &
        \multicolumn{7}{c}{\textbf{PAD-SMA}} \\
        \cmidrule(lr){2-15}
         & ASR$_{10}$ & ASR$_{20}$ & ASR$_{50}$ & ASR$_{100}$ & ASR$_{\infty}$ & $\bar{t}$ & VRAM
         & ASR$_{10}$ & ASR$_{20}$ & ASR$_{50}$ & ASR$_{100}$ & ASR$_{\infty}$ & $\bar{t}$ & VRAM \\
        \midrule
        
        \rowcolor{rowgray} BCA & 4.9 & 6.2 & 9.6 & 15.3 & 30.5 & 0.011 & \textbf{0.9} &
        2.6 & 4.2 & 13.2 & 15.8 & 16.3 & 0.074 & \textbf{0.9} \\
        
        BGA & 1.3 & 1.8 & 2.8 & 3.0 & 3.5 & 0.014 & 1.1 &
        1.5 & 1.5 & 1.5 & 1.5 & 1.5 & 0.088 & 1.2 \\
        
        \rowcolor{rowgray} Grosse & 4.9 & 6.2 & 9.5 & 15.0 & 28.8 & 0.012 & \textbf{0.9} &
        2.9 & 4.3 & 13.4 & 16.4 & 17.0 & 0.076 & \textbf{0.9} \\
        
        rFGSM & 0.8 & 0.9 & 2.5 & 9.2 & 62.9 & 0.011 & 1.0 &
        0.8 & 1.0 & 2.3 & 6.6 & 11.8 & 0.074 & 1.0 \\
        
        \rowcolor{rowgray} PGD-$\ell_1$ & 5.6 & 7.7 & 12.4 & 18.8 & 30.8 & 0.018 & 1.0 &
        2.3 & 5.2 & 14.4 & 18.7 & 20.6 & 0.101 & 1.0 \\
        
        PGD-$\ell_2$ & 0.8 & 0.8 & 3.8 & 12.7 & 69.2 & 0.012 & \textbf{0.9} &
        0.8 & 1.0 & 2.3 & 9.3 & 16.2 & 0.079 & 1.0 \\
        
        \rowcolor{rowgray} PGD-$\ell_\infty$ & 0.8 & 0.9 & 3.3 & 14.8 & 73.9 & 0.012 & \textbf{0.9} &
        0.8 & 0.9 & 2.6 & 9.4 & 15.3 & 0.076 & 1.0 \\
        
        iMaxMA & 0.8 & 0.9 & 2.6 & 11.8 & 86.9 & 0.052 & 1.1 &
        1.0 & 1.7 & 4.1 & 13.0 & 22.4 & 0.439 & 1.1 \\
        
        \rowcolor{rowgray} SMA & 9.2 & 11.4 & 17.6 & 29.9 & 35.4 & 0.044 & 1.1 &
        3.8 & 5.7 & 13.9 & 14.7 & 14.7 & 0.340 & 1.1 \\

        Mimicry & 8.8 & 18.2 & 37.8 & 57.5 & 66.1 & \textbf{0.011} & 1.2 &
        11.1 & 18.9 & 47.9 & 57.4 & 59.5 & \textbf{0.018} & 1.3 \\
        
        \rowcolor{rowgray} $\sigma$-zero & 10.2 & 14.5 & 30.0 & 42.2 & 45.0 & 0.027 & 1.0 & 
        25.8 & 49.2 & 65.4 & 68.2 & 68.3 & 0.151 & 1.2 \\
        
        CW & 7.7 & 12.4 & 22.9 & 34.1 & 41.6 & 0.105 & 2.2 &
        5.9 & 11.5 & 22.5 & 24.4 & 24.4 & 0.696 & 2.3 \\
        
        \rowcolor{rowgray} $\sigma$-binary & \textbf{21.9} & \textbf{30.5} & \textbf{62.5} & \textbf{85.7} & \textbf{96.6} & 0.068 & 1.1 &
        \textbf{29.2} & \textbf{57.4} & \textbf{84.5} & \textbf{92.2} & \textbf{93.9} & 0.448 & 1.3 \\
        \bottomrule
    \end{tabular}
    \label{tab:asr_under_defense}
\end{table*}

Table~\ref{tab:asr_under_defense} presents a comparative analysis of the proposed \(\sigma\)-binary attack against a wide range of state-of-the-art adversarial methods under four representative defense models: DNN, AT-rFGSM\(^k\), AT-MaxMA, and PAD-SMA.  
These models were selected as they represent the most robust and widely adopted defenses on the MalScan dataset—namely, AT-rFGSM\(^k\), AT-MaxMA, and PAD-SMA—alongside the standard DNN model as a baseline.  
Complete results for additional defense mechanisms are reported in Appendix~\ref{app:remaining_defenses}.

The table reports the \textbf{Attack Success Rate} (\(\text{ASR}_k\)) under multiple feature-modification budgets \(k \in \{10, 20, 50, 100, \infty\}\), where each \(k\) defines the Hamming distance limit \(d_H(\mathbf{x},\mathbf{x}^\star)\) between original and adversarial samples.  
We also include mean runtime (\(t_\text{mean}\)) and maximum GPU memory usage (VRAM) across all 3{,}104 malware samples to assess computational efficiency.  
All experiments follow the standardized setup in Section~\ref{subsection:experimental_setup} using identical hardware, software, and batch configurations to ensure fair and reproducible comparisons.

\subsubsection*{Effectiveness}

\paragraph{Consistent superiority across defenses.}
The proposed \(\sigma\)-binary attack achieves the highest \(\text{ASR}_k\) across all four defenses and modification budgets. On the baseline DNN, it reaches near-perfect evasion even under tight budgets (\(\text{ASR}_{10}=94.5\%\), saturating at \(100\%\) for \(k\geq50\)), outperforming strong baselines such as PGD-\(\ell_1\) and SMA.  
Under adversarially trained defenses, \(\sigma\)-binary maintains its advantage—achieving \(\text{ASR}_{50}=90.3\%\) and \(\text{ASR}_{100}=98.8\%\) against AT-rFGSM\(^k\), and \(\text{ASR}_{100}=85.7\%\) under AT-MaxMA—while sustaining robust performance (\(\text{ASR}_{100}=92.2\%\)) even against the detector-based PAD-SMA.

\paragraph{Robustness to binarization.}
Attacks originally optimized in continuous space (e.g., CW, \(\sigma\)-zero) show sharp degradation after rounding, as their fine-grained perturbations near decision boundaries are disrupted by binary projection.  
In contrast, \(\sigma\)-binary’s discrete-aware optimization and prioritized rounding maintain stability across defenses, producing consistently higher ASR values and confirming the importance of binary-native optimization for malware domains.

\paragraph{Strengthened Mimicry remains non-competitive.}
To strengthen comparison, the Mimicry attack was extended to use all benign test samples rather than the 30 random references (Mimicry\(\times30\)) used previously~\cite{li2020adversarial,li2023pad}.  
Although Mimicry\(\times30\) was among the strongest attacks in prior work, this enhanced version still performs substantially worse, reaching only \(\text{ASR}_{\infty}=66.1\%\) under AT-MaxMA and \(59.5\%\) under PAD-SMA—far below \(\sigma\)-binary’s \(>90\%\) success.  
This confirms that feature-level resemblance to benign software is insufficient to bypass modern adversarial defenses.

\paragraph{Stable performance under tight budgets.}
While competing methods show inconsistent trends across defenses and budgets, \(\sigma\)-binary maintains stable high performance even with minimal feature modifications (\(k=10\)), highlighting the efficiency of its distance-aware optimization and rounding strategy.

\begin{figure*}[t]
    \centering
    \includegraphics[width=\linewidth]{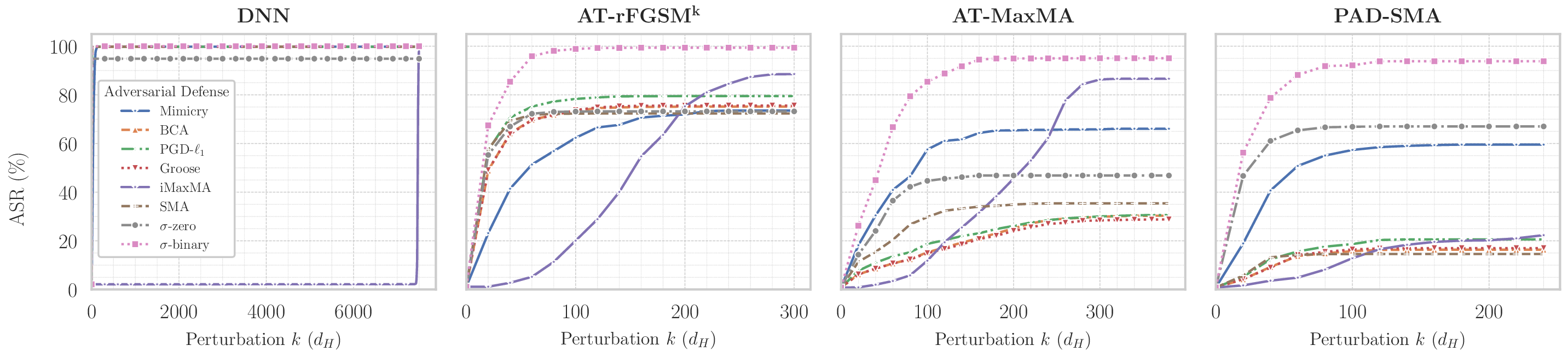}
    \caption{Attack Success Rate (ASR) versus perturbation budget \(k\) (Hamming distance \(d_H\)) across four defense models.
    Only the most competitive attacks are shown for clarity.}
    \label{fig:asr_vs_k}
\end{figure*}

\subsubsection*{Reliability and Convergence Behavior}
Complementary to Table~\ref{tab:asr_under_defense}, we present the robustness evaluation curves in Figure~\ref{fig:asr_vs_k} for each attack under four representative defenses. Similar curves for all remaining defense models are included in Appendix~\ref{app:remaining_defenses}. Due to the large number of evaluated attacks, only the most competitive methods are shown in Figure~\ref{fig:asr_vs_k} for clarity, ensuring readability while preserving the key comparative trends across defenses.
Across all defenses, \(\sigma\)-binary exhibits a steep and stable growth curve, saturating near \(100\%\) for \(k<100\), while alternatives such as SMA and Mimicry show slower or irregular improvements.  
Under the strongest defenses (PAD-SMA and AT-MaxMA), \(\sigma\)-binary consistently outperforms competitors by 20–30 percentage points, demonstrating stable convergence and high reliability across perturbation ranges.  
Unlike gradient-based baselines that often saturate early or fail to converge fully, \(\sigma\)-binary almost always finds valid adversarial solutions, ensuring faithful robustness evaluation.

\subsubsection*{Efficiency}
Despite its stronger optimization framework, \(\sigma\)-binary remains computationally practical.  
Its mean runtime is moderately higher than lightweight attacks due to nested optimization and adaptive rounding but remains computationally efficient overall.
For instance, under PAD-SMA, the proposed \(\sigma\)-binary attack completes inference with a mean runtime of approximately \(0.45~\text{s/sample}\) while using only \(1.3~\text{GB}\) of VRAM—demonstrating practical computational efficiency even on a standard mobile workstation. 
This balance between effectiveness and efficiency demonstrates that \(\sigma\)-binary’s robustness does not come at the cost of prohibitive resource requirements.

\subsubsection*{Ablation Study on \texorpdfstring{\(\sigma\)}{sigma}-binary}
\label{subsubsection:ablation_sigma_binary}

To further investigate the contribution of each component in the proposed \(\sigma\)-binary framework, 
we conducted an ablation study under two representative adversarially trained defenses: 
AT-rFGSM\(^k\) and AT-MaxMA. 
Each variant disables one major component of the framework while keeping all other parameters fixed, 
allowing an isolated assessment of its impact on attack performance. 
The examined components include: 
(i) the differentiable Hamming distance (${d}_{H}$) approximation, 
(ii) the gradient normalization factor, 
(iii) the adaptive sparsity projection operator, 
(iv) the Prioritized Binary Rounding (PBR) mechanism, and 
(v) the dynamic confidence adjustment strategy. 
Performance is measured using attack success rates \(\text{ASR}_k\) at modification budgets \(k \in \{10, 20, 50, 100, \infty\}\), 
as well as the median and interquartile range (IQR) of perturbation magnitudes \(\tilde{d}_H\) among successful attacks. 
To ensure stability and comparability, all experiments in this ablation were performed on a fixed subset of 1{,}000 randomly sampled test instances with an extended optimization budget of 10{,}000 iterations. 
The detailed results are reported in Table~\ref{tab:ablation_sigma_binary}.

\begin{table*}[t!]
\centering
\caption{Ablation study of the proposed \(\sigma\)-binary attack under two adversarially trained defenses (AT-rFGSM\(^k\) and AT-MaxMA).  
Each variant removes one core component while keeping all others unchanged.  
\(\text{ASR}_k\) denotes the attack success rate (\%) under a Hamming distance constraint \(d_H(\mathbf{x},\mathbf{x}^\star)\leq k\).  
\(\tilde{d}_{H,\text{median}}\) and IQR represent the median and interquartile range of perturbation magnitudes among successful attacks.  
Bold values indicate the best performance within each defense.}
\label{tab:ablation_sigma_binary}
\vspace{2mm}
\small
\setlength{\tabcolsep}{2pt}
\renewcommand{\arraystretch}{1.1}
\begin{tabular}{lcccccc|cccccc}
\toprule
\multirow{2}{*}{\textbf{Configuration}} &
\multicolumn{6}{c|}{\textbf{AT-rFGSM\(^k\)}} &
\multicolumn{6}{c}{\textbf{AT-MaxMA}} \\
\cmidrule(lr){2-7} \cmidrule(lr){8-13}
 & ASR\(_{10}\) & ASR\(_{20}\) & ASR\(_{50}\) & ASR\(_{100}\) & ASR\(_{\infty}\) & Med / IQR
 & ASR\(_{10}\) & ASR\(_{20}\) & ASR\(_{50}\) & ASR\(_{100}\) & ASR\(_{\infty}\) & Med / IQR \\
\midrule
\rowcolor{rowgray} Full \(\sigma\)-binary (all) &
\textbf{51.1} & \textbf{69.0} & \textbf{91.9} & \textbf{98.9} & \textbf{99.4} & \textbf{10.0 / 19.0} &
\textbf{23.8} & \textbf{36.9} & 64.3 & \textbf{86.9} & \textbf{97.0} & \textbf{31.0 / 53.0} \\
\midrule
Without ${d}_{H}$  approximation &
7.5 & 19.8 & 40.3 & 75.0 & 99.3 & 68.0 / 73.0 &
1.8 & 3.2 & 12.5 & 17.5 & 94.9 & 215.0 / 116.0 \\
\rowcolor{rowgray} Without normalization &
45.0 & 63.1 & 87.6 & 98.2 & 99.1 & 12.0 / 26.0 &
18.7 & 25.8 & 59.5 & 85.6 & 96.9 & 42.0 / 52.0 \\
Without adaptive projection &
50.4 & 68.2 & 91.2 & \textbf{98.9} & \textbf{99.4} & \textbf{10.0} / 21.0 &
23.1 & 36.6 & 64.2 & 86.7 & \textbf{97.0} & 33.0 / 55.0 \\
\rowcolor{rowgray} Without PBR rounding &
50.4 & 68.8 & 91.8 & \textbf{98.9} & \textbf{99.4} & \textbf{10.0} / 20.0 &
23.0 & 36.0 & 63.1 & 86.7 & \textbf{97.0} & 32.0 / 54.0 \\
Without dynamic confidence &
50.0 & 68.5 & \textbf{91.9} & \textbf{98.9} & \textbf{99.4} & \textbf{10.0} / 20.0 &
23.3 & 34.8 & \textbf{64.6} & \textbf{86.9} & \textbf{97.0} & 35.0 / 54.5 \\
\bottomrule
\end{tabular}
\end{table*}

As shown in Table~\ref{tab:ablation_sigma_binary}, each component contributes meaningfully to the performance of the \(\sigma\)-binary framework.  
Removing the differentiable ${d}_{H}$ approximation causes the largest drop in low-budget success rates (\(k \leq 50\)) and a substantial increase in perturbation distance, confirming its key role in stabilizing gradient-based optimization.
Disabling gradient normalization factor or adaptive projection slightly reduces \(\text{ASR}_k\) and increases variability, indicating that both components help maintain balanced updates and sparsity.  
When the PBR mechanism is replaced with deterministic binary rounding, the results remain relatively strong—thanks to checkpointing of the best rounded candidates—but perturbations become less sparse and slightly larger.  
Finally, omitting dynamic confidence adjustment yields comparable final ASR values but broader perturbation dispersion (higher IQR), suggesting less consistent convergence.  
Overall, the results confirm that the combination of smooth distance modeling, adaptive sparsity, and prioritized rounding is crucial for achieving the high efficiency and reliability of the full \(\sigma\)-binary attack.

\subsubsection*{Ablation Study on PBR}
\label{subsubsection:ablation-pbr}

To quantify the contribution of the \textit{feature-importance weighting} mechanism in the proposed PBR framework, we performed an ablation study comparing two variants:  
(i) a baseline configuration that prioritizes features solely according to perturbation magnitude (derived from gradient magnitude), and  
(ii) an enhanced configuration that additionally incorporates discriminative feature importance derived from the training data.  

In the enhanced version, discriminative feature importance acts as a secondary prioritization criterion: when multiple features share the same gradient magnitude, their order is resolved according to discriminative importance.  
Consequently, its overall impact remains modest, but it can provide incremental improvements in cases where gradient information alone is insufficient to distinguish between features of equal significance.

\begin{table*}[t!]
\centering
\caption{Ablation study on the effect of incorporating feature importance in PBR. \(\text{ASR}_k\) denotes the attack success rate (\%) under a Hamming distance constraint \(d_H(\mathbf{x},\mathbf{x}^\star)\leq k\). \(\tilde{d}_{H,\mathrm{median}}\) and IQR represent the median and interquartile range of normalized Hamming distances across successful attacks.}
\label{tab:ablation-pbr-redesigned}

\setlength{\tabcolsep}{7pt}
\renewcommand{\arraystretch}{1.15}

\begin{tabular}{@{} l c l
S[table-format=2.2]
S[table-format=2.2]
S[table-format=2.2]
S[table-format=2.2]
S[table-format=2.2]
S[table-format=2.2]
S[table-format=2.2]
S[table-format=2.2] @{}}
\toprule
\textbf{Model} & \textbf{Iter.} & \textbf{Method} &
{\( \text{ASR}_{10} \)} &
{\( \text{ASR}_{20} \)} &
{\( \text{ASR}_{50} \)} &
{\( \text{ASR}_{100} \)} &
{\( \text{ASR}_\infty \)} &
{\( \tilde{d}_{H,\mathrm{median}} \)} &
{\textbf{IQR}} \\
\midrule

\multirow{4}{*}{\textbf{AT-rFGSM\(^k\)}} 
& \multirow{2}{*}{100} & Perturbation Magnitude & 32.02 & 49.48 & 74.48 & 90.43 & 92.20 & 18.00 & 33.00 \\
&  & + Feature Importance & \textbf{32.15} & \textbf{49.55} & \textbf{74.52} & \textbf{90.46} & 92.20 & 18.00 & 33.00 \\
\cmidrule(lr){2-10}
& \multirow{2}{*}{1000} & Perturbation Magnitude & 47.94 & 66.95 & 90.30 & 98.81 & 99.39 & 11.00 & 22.00 \\
&  & + Feature Importance & 47.94 & 66.95 & 90.30 & 98.81 & 99.39 & 11.00 & 22.00 \\
\midrule

\multirow{4}{*}{\textbf{AT-MaxMA}}
& \multirow{2}{*}{100} & Perturbation Magnitude & 15.79 & 22.97 & 53.16 & 84.54 & 95.52 & 47.00 & 55.0 \\
&  & + Feature Importance & \textbf{15.88} & \textbf{23.10} & 53.16 & 84.54 & 95.52 & 47.00 & 51.00 \\
\cmidrule(lr){2-10}
& \multirow{2}{*}{1000} & Perturbation Magnitude & 21.97 & 30.48 & 62.50 & 85.76 & 96.59 & 38.00 & 54.00 \\
&  & + Feature Importance & 21.97 & 30.48 & 62.50 & 85.76 & 96.59 & 38.00 & 54.00 \\

\bottomrule
\end{tabular}
\end{table*}

Both variants share identical hyperparameter settings and are evaluated on a fixed subset of 1{,}000 randomly sampled test instances to ensure fairness.  
Each experiment was run for 100 and 1{,}000 optimization iterations to examine both early-stage and converged behavior.  
Table~\ref{tab:ablation-pbr-redesigned} reports the attack success rates (\(\text{ASR}_k\)) under different perturbation budgets \(k \in \{10, 20, 50, 100, \infty\}\), together with the median and interquartile range (IQR) of normalized Hamming distances \((\tilde{d}_{H,\mathrm{median}}, \mathrm{IQR})\) among successful attacks.

At smaller iteration budgets (\(100\)), incorporating discriminative feature importance yields consistent though marginal gains of approximately \textbf{0.1--0.2 percentage points} in \(\text{ASR}_k\) across most perturbation thresholds while maintaining nearly identical distance distributions.  
This indicates that secondary ordering by discriminative importance helps guide the rounding process toward semantically meaningful binary features, improving optimization efficiency in some instances.

When the iteration budget increases to \(1{,}000\), both variants converge to virtually identical performance.  
The ASR, median distances, and IQR values become indistinguishable, suggesting that with sufficient optimization steps, the attack inherently learns to emphasize salient features even without explicit feature-importance guidance.  

Overall, these findings show that incorporating discriminative feature importance as a secondary tie-breaking criterion can slightly improve early-stage optimization but has negligible influence after full convergence, thereby confirming that PBR’s primary effectiveness stems from gradient-based prioritization.

\subsubsection*{Hyperparameter Robustness of the \texorpdfstring{\(\sigma\)}{sigma}-Binary Attack}
\label{subsubsection:hyperparam-robustness}

We examined the sensitivity of the proposed \(\sigma\)-binary attack to its key hyperparameters: the smoothing factor \(\sigma\), the initial sparsity threshold \(\gamma_0\), and the sparsity adjustment rate \(t\).  
All experiments were performed on 1{,}000 randomly selected test samples for both AT-rFGSM\(^k\) and AT-MaxMA models.

For the ablation on the sparsity adaptation rate \(t\), the other hyperparameters were fixed to their default values of \(\gamma_0 = 0.3\) and \(\sigma = 10^{-3}\).  
In contrast, in the preceding experiments examining the effects of \(\gamma_0\) and \(\sigma\), these parameters were systematically varied within the ranges \(\gamma_0 \in [0, 0.5]\) and \(\sigma \in [10^{-6}, 10^{-1}]\), respectively, while the sparsity adaptation rate was held constant at its default value of \(t = 0.01\).  
For each configuration, we report the Attack Success Rate (ASR) at multiple perturbation budgets and the median normalized Hamming distance (\(\tilde{d}_{H,\mathrm{median}}\)).

\begin{figure*}[t!]
  \centering
  \includegraphics[width=\textwidth]{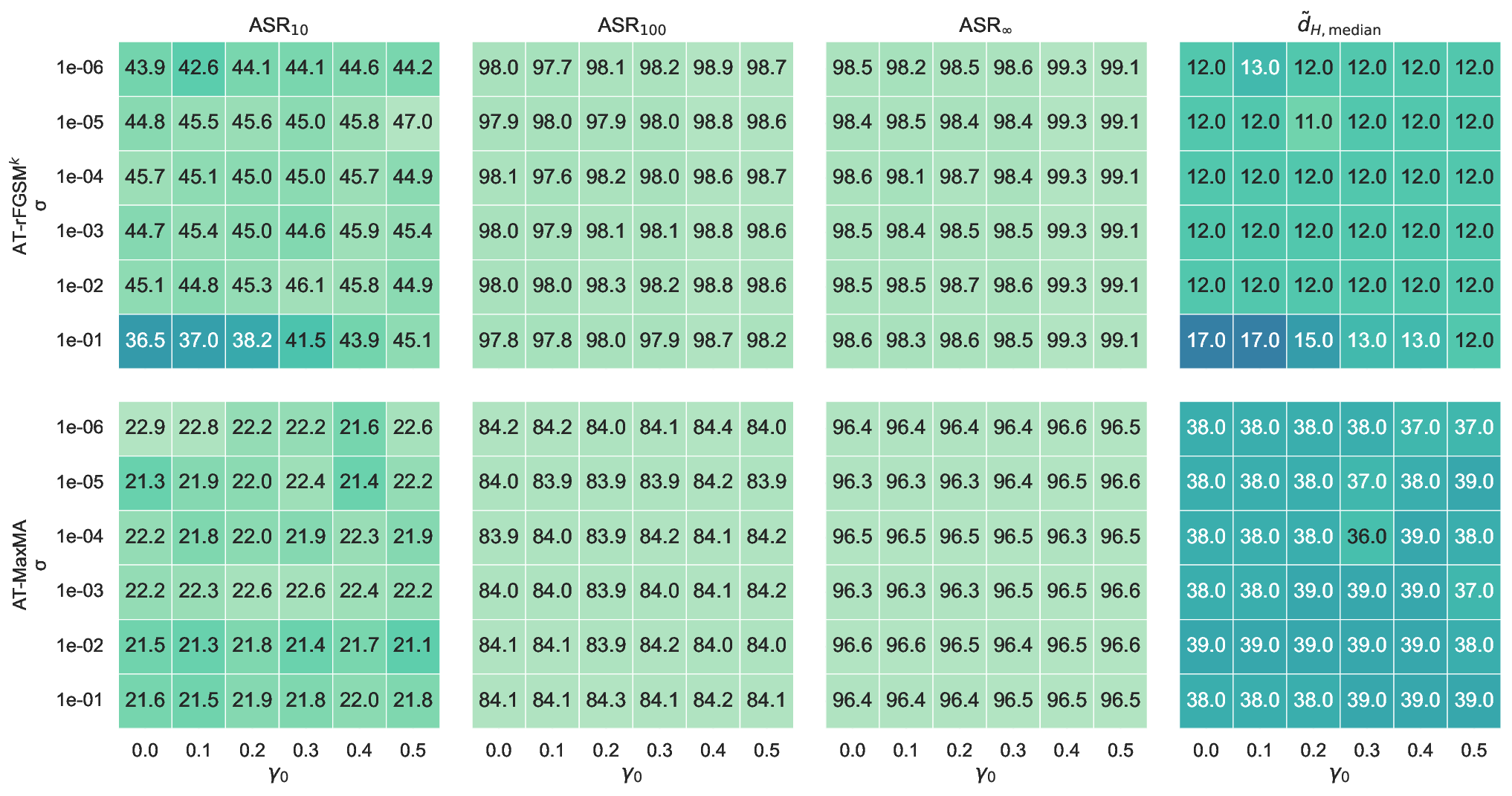}
  \caption{Hyperparameter robustness of the \(\sigma\)-binary attack across smoothing parameter \(\sigma\) and initial sparsity threshold \(\gamma_0\).
Each cell shows the Attack Success Rate (ASR) at different perturbation budgets and the corresponding median normalized Hamming distance.
Top panel: AT-rFGSM\(^k\); bottom panel: AT-MaxMA.}
  \label{fig:heatmaps}
\end{figure*}

\textbf{Effect of \(\gamma_0\) and \(\sigma\):}
Fig.~\ref{fig:heatmaps} summarizes the interaction between the smoothing parameter \(\sigma\) and the initial sparsity threshold \(\gamma_0\).  
Across all tested values, the attack performance remains remarkably stable.  
Both AT-rFGSM\(^k\) and AT-MaxMA exhibit consistent ASR and \(\tilde{d}_{H,\mathrm{median}}\), confirming that the adaptive projection operator \(\Pi_{\gamma}\) rapidly neutralizes the impact of \(\gamma_0\).  
Similarly, moderate values of \(\sigma\) (between \(10^{-6}\) and \(10^{-2}\)) yield comparable results.  
Only very large \(\sigma\) values (\(10^{-1}\)) slightly degrade ASR at small perturbation budgets and increase \(\tilde{d}_{H,\mathrm{median}}\), due to excessive smoothing of the Hamming surrogate that reduces sparsity in early optimization steps.

\begin{figure*}[t!]
  \centering
  \includegraphics[width=\textwidth]{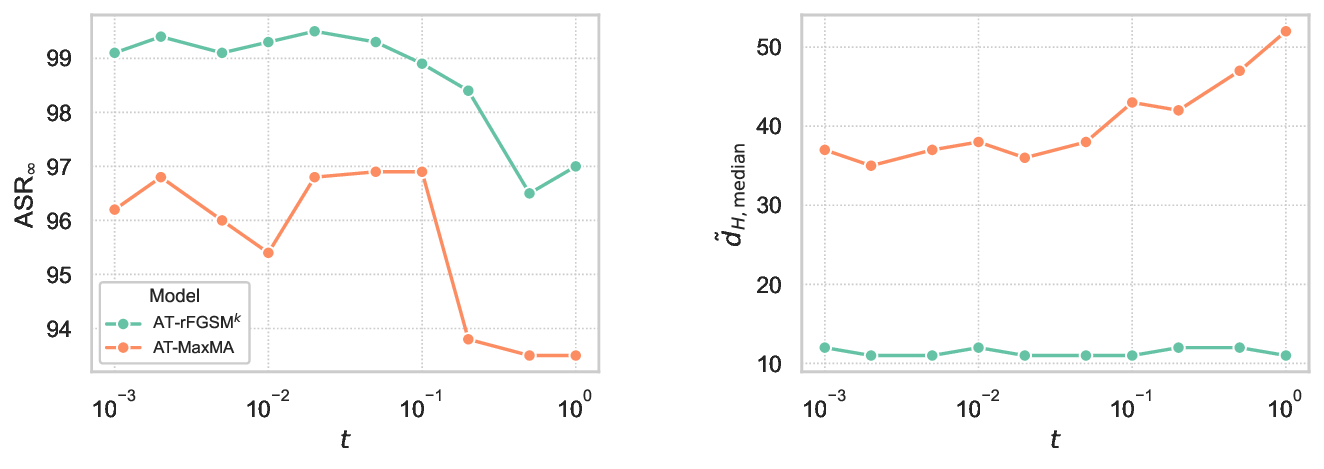}
  \caption{Effect of the sparsity adjustment factor \(t\) on the \(\sigma\)-binary attack.
Left: ASR\(_\infty\) versus \(t\) (log scale). Right: median normalized Hamming distance \(\tilde{d}_{H,\mathrm{median}}\) versus \(t\).}
  \label{fig:tuning_t}
\end{figure*}

\textbf{Effect of the sparsity adjustment factor \(t\):}
Fig.~\ref{fig:tuning_t} illustrates the impact of the sparsity update rate \(t\).  
The \(\sigma\)-binary attack remains robust for \(t \leq 10^{-1}\), maintaining high ASR and low \(\tilde{d}_{H,\mathrm{median}}\) across both defenses.  
For larger \(t\), AT-MaxMA becomes more sensitive, showing a drop in ASR\(_\infty\) and an increase in \(\tilde{d}_{H,\mathrm{median}}\).  
This behavior indicates that overly aggressive updates of \(\gamma\) can destabilize the optimization process, leading to less compact perturbations.

\textbf{Note on dynamic confidence adjustment:}
We did not include a separate sensitivity analysis for the dynamic confidence adjustment parameters, as this mechanism becomes active only after convergence of the loss function. 
Its role is to refine already converged adversarial examples by slightly perturbing the confidence margins around the decision boundary, without altering the loss or optimization trajectory.  
As shown in Appendix~\ref{tab:ablation_sigma_binary}, the effect of this mechanism is marginal—primarily improving results under high iteration budgets and stronger defenses by stabilizing late-stage convergence.  
Because it does not materially influence the main optimization trajectory, its hyperparameter tuning has negligible practical impact and is therefore excluded from this robustness study.

\textbf{Discussion:}
Overall, the \(\sigma\)-binary attack demonstrates strong robustness to its principal hyperparameters.  
A broad configuration range produces near-optimal performance, enabling efficient deployment without per-instance tuning.  
Empirically, we recommend the following default ranges:
\[
\sigma \in [10^{-6},10^{-2}], \quad t \leq 10^{-1}, \quad \gamma_0 \in [0,0.5].
\]
These settings consistently yield high ASR with minimal perturbation size for both evaluated defenses.

\textbf{Answer to RQ1: }\(\sigma\)-binary achieves consistently higher attack success rates across all defenses while maintaining practical efficiency and remarkable stability.  
Its advantage stems from the integrated design of differentiable distance modeling, adaptive sparsity control, and prioritized rounding, which together enable precise and reliable optimization in binary feature spaces.  
The attack remains robust to hyperparameter variation and maintains convergence stability even under strong adversarial training, providing both superior effectiveness and trustworthy evaluation of malware defense robustness.

\subsection{RQ2: Baseline performance of defenses}

This subsection evaluates the performance of defense models in the absence of adversarial attacks. The defenses are categorized into two groups: (1) those without adversary detectors (DNN, AT-rFGSM\(^k\), AT-MaxMA) and (2) those with adversary detectors (DNN\textsuperscript{+}, KDE, DLA, ICNN, PAD-SMA), operating in either \textit{deferred} or \textit{conservative} modes.

For clarity and consistency, we refer to \textbf{conservative-mode models} simply by their base names (e.g., DLA, KDE, ICNN), while the \textbf{deferred-mode variants} are explicitly indicated with the suffix “(Deferred)” throughout the paper.

As summarized in Table~\ref{tab:effectiveness_absence_attacks}, models with adversary detectors in deferred mode consistently outperform their conservative counterparts. Deferred models exclude a small percentage of challenging samples, which simplifies classification and boosts accuracy. Notably, DNN\textsuperscript{+} (Deferred) achieves the highest accuracy (98.1\%) and F1 score (98.1\%), excluding 5.0\% of samples. Similarly, DLA (Deferred) and KDE (Deferred) achieve competitive results with F1 scores of 97.9\% and 97.3\%, respectively, while excluding 5.5\% and 3.7\% of samples.

\definecolor{rowgray}{gray}{0.95}

\begin{table}[tbp]
    \centering
    \caption{Effectiveness (\%) of detectors in the absence of adversarial attacks.}
    \label{tab:effectiveness_absence_attacks}
    \setlength{\tabcolsep}{6pt}
    \begin{tabular}{lccccc}
        \toprule
        \textbf{Defense} & \textbf{FNR} & \textbf{FPR} & \textbf{Acc} & \textbf{F1} & \textbf{ExS} \\
        \midrule
        DNN & 2.1 & 3.1 & 97.4 & 97.4 & - \\
        \rowcolor{rowgray} AT-rFGSM\(^k\) & 1.1 & 6.5 & 96.2 & 96.4 & - \\
        AT-MaxMA & \textbf{0.8} & 9.6 & 94.9 & 95.1 & - \\
        \rowcolor{rowgray} KDE (Deferred) & 2.2 & 3.2 & 97.3 & 97.3 & 3.7 \\
        KDE & 2.1 & 5.5 & 96.2 & 96.3 & - \\
        \rowcolor{rowgray} DLA (Deferred) & 1.5 & 2.6 & 97.9 & 97.9 & 5.5 \\
        DLA & 1.4 & 4.3 & 97.2 & 97.3 & - \\
        \rowcolor{rowgray} DNN\textsuperscript{+} (Deferred) & 1.5 & \textbf{2.2} & \textbf{98.1} & \textbf{98.1} & 5.0 \\
        DNN\textsuperscript{+} & 1.5 & 6.2 & 96.2 & 96.3 & - \\
        \rowcolor{rowgray} ICNN (Deferred) & 2.1 & 3.2 & 97.4 & 97.5 & 5.1 \\
        ICNN & 2.0 & 10.6 & 93.7 & 94.1 & - \\
        \rowcolor{rowgray} PAD-SMA (Deferred) & 0.9 & 7.9 & 95.6 & 95.7 & 4.6 \\
        PAD-SMA & 0.8 & 10.4 & 94.4 & 94.7 & - \\
        \bottomrule
    \end{tabular}
\end{table}

Conservative-mode models prioritize strict classification, treating all suspicious samples flagged by the detector \(g\) as malicious. This approach results in elevated false positive rates (FPRs), reducing overall performance. For instance, ICNN has an FPR of 10.6\%, compared to 3.2\% in its deferred counterpart. Similarly, PAD-SMA shows an accuracy of 94.4\%, lower than its deferred counterpart (95.6\%).

Adversarially hardened models, such as AT-rFGSM\(^k\), \mbox{AT-MaxMA}, and PAD-SMA, exhibit lower false negative rates (FNRs), highlighting their ability to detect malicious samples effectively. However, this robustness against adversarial attacks comes at the cost of performance under benign conditions. For example, AT-rFGSM\(^k\) achieves an FNR of 1.1\% and an FPR of 6.5\%, resulting in an accuracy of 96.2\%. PAD-SMA (Deferred) achieves an FNR of 0.9\% and an FPR of 7.9\%, with an accuracy of 95.6\%, lagging behind other deferred models due to its reliance on adversarial training, which biases its detector \(g\) and increases FPR \cite{li2023pad}.

\textbf{Answer to RQ2: }In the absence of attacks, deferred-mode defenses such as DNN\textsuperscript{+}~(Deferred), DLA~(Deferred), and KDE~(Deferred) achieve superior accuracy and F1 scores by excluding challenging samples, whereas conservative-mode models and adversarially hardened defenses face trade-offs between robustness and performance, with higher FPRs and lower accuracy when there are no adversarial attacks.

\subsection{RQ3: Robustness against oblivious attacks}
\label{subsection:Robustness_oblivious_attacks}

In this subsection, we assess the robustness of five defenses—KDE, DLA, DNN\textsuperscript{+}, ICNN, and PAD-SMA—against oblivious attacks using the proposed \textit{\(\sigma\)-binary} attack. These defenses integrate an adversary detector \(g\), which is not explicitly targeted in this evaluation.

Oblivious attacks operate under the assumption that attackers are unaware of the presence or functionality of the detector \(g\), focusing solely on evading the main classifier. This scenario provides a baseline for evaluating the intrinsic robustness of \(g\) and the overall defense system.

While non-adaptive attack evaluations are necessary, they are insufficient to fully gauge robustness. Success against these weaker adversaries does not guarantee resilience to adaptive attacks~\cite{carlini2019evaluating}. However, these evaluations can still highlight vulnerabilities in defenses under minimal adversarial threats.

The robustness evaluation considers two configurations: \textit{deferred} and \textit{conservative}. In the deferred mode, samples flagged as suspicious by \(g\) are excluded from the ASR calculation, isolating the vulnerabilities of the main classifier. In contrast, the conservative mode assumes all suspicious samples flagged by \(g\) are malicious, incorporating them into the ASR calculation. Unless explicitly labeled as \textit{Deferred}, the results represent the conservative configuration.

\begin{figure}[tbp]
    \centering
    \includegraphics[width=0.8\linewidth]{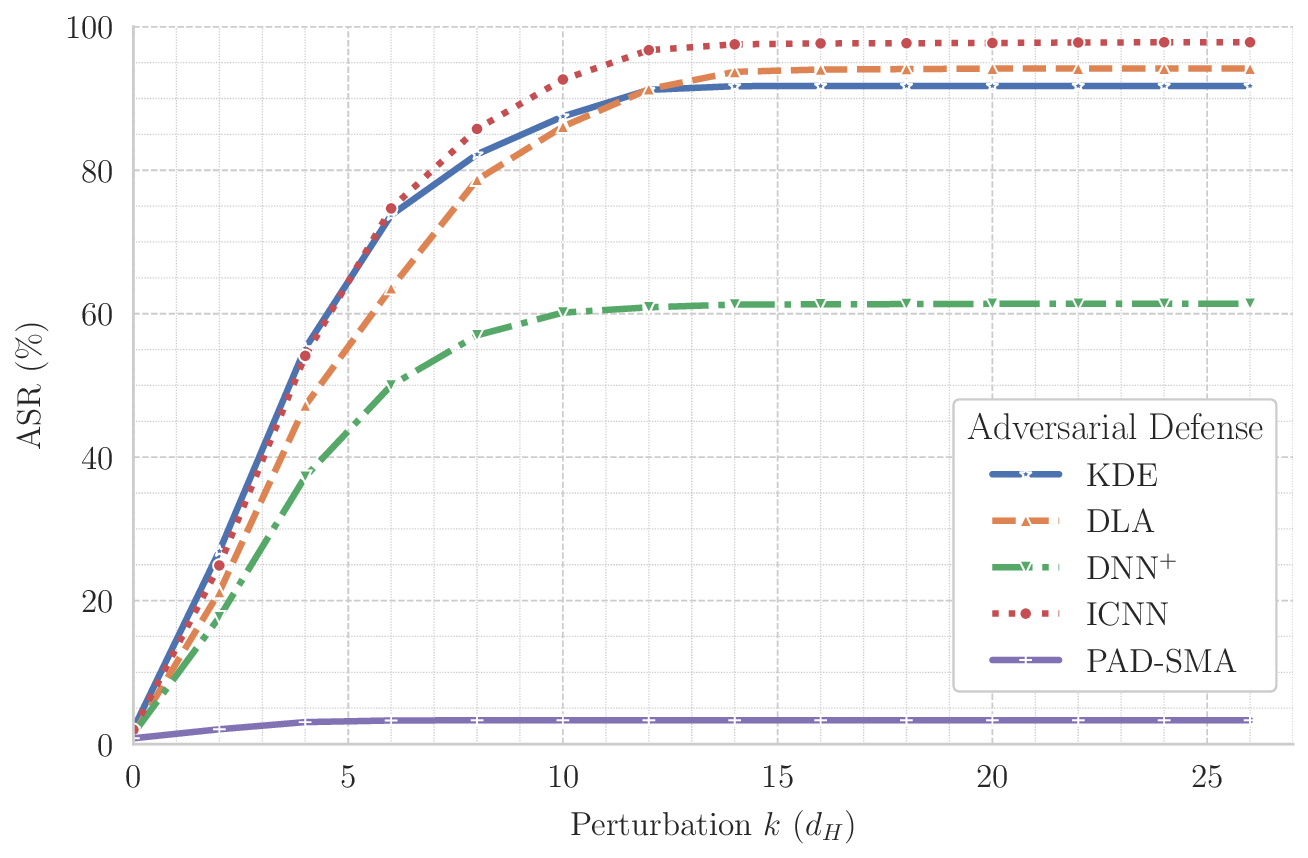}
    \caption{Robustness evaluation curves (ASR vs. perturbation budget \(k\)) against oblivious attacks.}
    \label{fig:robustness_oblivious}
\end{figure}

\definecolor{rowgray}{gray}{0.95}

\begin{table}[tbp]
    \centering
    \caption{Robustness against Oblivious Attacks}
    \label{tab:robustness_oblivious}
    \setlength{\tabcolsep}{5pt} 
    \begin{tabular}{lcccc}
        \toprule
        \textbf{Defense} & \textbf{\( \text{ASR}_\infty \) } & \textbf{ExS} & \(\boldsymbol{\tilde{d}_{H,\mathrm{median}}}\) & \textbf{\( \text{IQR} \)}\\
        \midrule
        \rowcolor{rowgray} KDE (Deferred) & 100.0 & 8.3 & 4.0 & 4.0 \\
        KDE & 91.8 & - & 4.0 & 4.0 \\
        \rowcolor{rowgray} DLA (Deferred) & 100.0 & 5.8 & 4.0 & 4.0 \\
        DLA & 94.2 & - & 4.0 & 4.0 \\
        \rowcolor{rowgray} DNN\textsuperscript{+} (Deferred) & 100.0 & 38.6 & 4.0 & 4.0\\
        DNN\textsuperscript{+} & 61.4 & - & 4.0 & 4.0 \\
        \rowcolor{rowgray} ICNN (Deferred) & 100.0 & 2.1 & 4.0 & 4.0 \\
        ICNN & 97.9 & - & 4.0 & 4.0 \\
        \rowcolor{rowgray} PAD-SMA (Deferred) & 100.0 & 96.7 & 2.0 & 2.0 \\
        PAD-SMA & 3.3 & - & 2.0 & 2.0 \\
        \bottomrule
    \end{tabular}
\end{table}

Table~\ref{tab:robustness_oblivious} summarizes the evaluation results. In the deferred mode, all defenses exhibit \( \text{ASR}_{\infty} \) of 100\%, underscoring the inability of the base classifier to withstand attacks in isolation. However, the conservative mode reveals substantial variations in robustness, highlighting the critical role of the detector \(g\) in mitigating adversarial threats.

Among the evaluated defenses, PAD-SMA exhibits exceptional robustness, achieving an ASR of just 3.3\% in its conservative configuration. The flat ASR curve, shown in Figure~\ref{fig:robustness_oblivious} highlights its consistent resilience across varying perturbation budgets. This superior performance, compared to ICNN—which shares a similar defense architecture—can be attributed to PAD-SMA's incorporation of adversarial training, which significantly enhances the robustness of the adversary detector \(g\).

ICNN exhibits a high ASR of 97.9\%, and its steep ASR curve, as depicted in Figure~\ref{fig:robustness_oblivious}, underscoring substantial vulnerability to small perturbations. Similarly, KDE and DLA perform poorly, with ASRs of 91.8\% and 94.2\%, respectively, in the conservative configuration. DNN\textsuperscript{+} shows moderate improvement, achieving an ASR of 61.4\%, yet it remains significantly less robust than PAD-SMA in resisting oblivious attacks.

\textbf{Answer to RQ3:} PAD-SMA demonstrates unparalleled robustness against oblivious attacks, outperforming other defenses by a wide margin. The considerable vulnerability of ICNN, KDE, and DLA emphasizes the indispensable role of adversarial training in crafting effective defense mechanisms.

\subsection{RQ4: Robustness Against Adaptive Attacks}
\label{subsec:Adaptive Attacks}

This subsection evaluates the robustness of defenses against adaptive attacks using our proposed \textit{\(\sigma\)-binary} attack. Unlike oblivious attacks, adaptive attacks assume full knowledge of the adversary detector \(g\), enabling attackers to craft perturbations that evade both the primary model and \(g\), presenting a greater challenge for defenses.

\begin{figure}[tbp]
    \centering
    \includegraphics[width=0.8\linewidth]{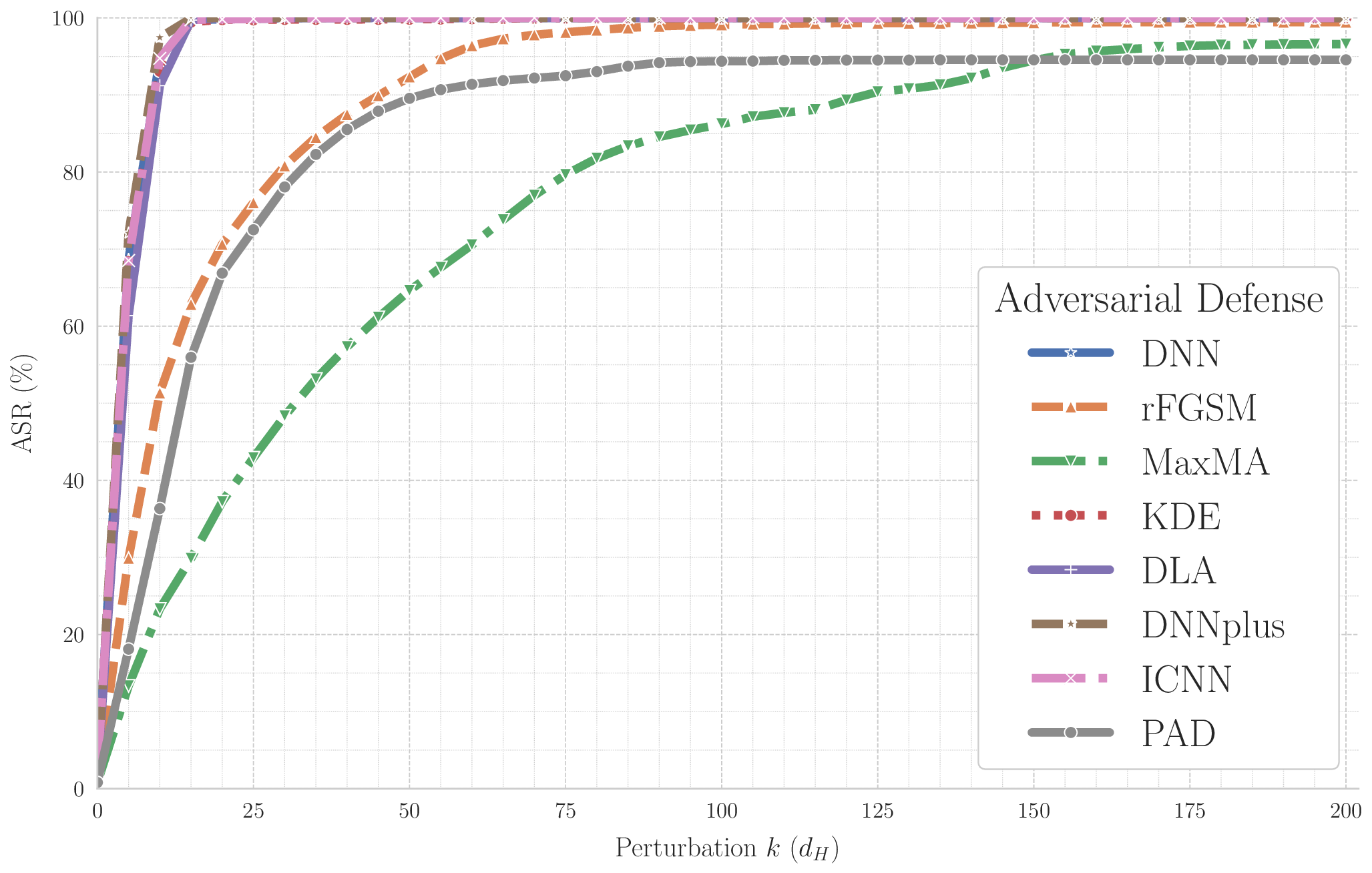}
    \caption{Robustness evaluation curves (ASR vs. perturbation budget \(k\)) against adaptive attacks.}
    \label{fig:robustness_adaptive}
\end{figure}

Table~\ref{tab:effectiveness_adaptive} and Figure~\ref{fig:robustness_adaptive} present the robustness of each defense under adaptive attacks. DNN, KDE, DLA, DNN\textsuperscript{+}, and ICNN exhibit high vulnerability, with a median perturbation~(\( \tilde{d}_{H,\mathrm{median}} \)) of 4.0 and \( \text{ASR}_{10} \) exceeding 90\%, indicating susceptibility to small perturbations. 

In contrast, adversarially trained defenses such as \mbox{PAD-SMA}, AT-MaxMA, and AT-rFGSM\(^k\) show significantly enhanced robustness. AT-MaxMA achieves the strongest resistance at low perturbation budgets, with \( \text{ASR}_{10} = 23.3\%\), outperforming PAD-SMA (36.3\%) and AT-rFGSM\(^k\) (51.4\%). This highlights AT-MaxMA’s capability to mitigate low-budget attacks effectively.

At moderate perturbation budgets, as indicated by \( \text{ASR}_{50} \), AT-MaxMA maintains its superiority, achieving the lowest value (65.2\%), followed by PAD-SMA (89.8\%) and \mbox{AT-rFGSM\(^k\)} (92.0\%). However, for high perturbation budgets, PAD-SMA demonstrates greater robustness, achieving the lowest \( \text{ASR}_{\infty} = 94.6\%\), slightly outperforming \mbox{AT-MaxMA} (96.6\%) and other defenses. PAD-SMA’s performance reflects its ability to restrict the success of high-budget attacks, as evidenced by its flattened \( \text{ASR}_k \) curve for \(k \geq 50\).

These findings highlight a clear trade-off between robustness at different perturbation budgets and practical deployment factors. 
AT-MaxMA is particularly effective against small perturbations, exhibiting superior robustness at low budgets while maintaining lower false-positive rates (FPR) and higher benign accuracy. 
Conversely, PAD-SMA demonstrates stronger resistance under high-budget attacks (\(k \geq 50\)) but at the cost of higher FPR. 
Its consistent performance in both oblivious and adaptive settings (see Table~\ref{tab:robustness_oblivious}) underscores the benefits of adversarial training for comprehensive defense coverage.

\definecolor{rowgray}{gray}{0.95}

\begin{table}[tbp]
    \centering
    \caption{Robustness against Adaptive Attacks}
    \label{tab:effectiveness_adaptive}
    \footnotesize 
    \setlength{\tabcolsep}{5pt} 
    \begin{tabular}{lccccc}
        \toprule
        \textbf{Defense} &
        \textbf{\( \text{ASR}_{10} \)} &
        \textbf{\( \text{ASR}_{50} \)} &
        \textbf{\( \text{ASR}_\infty \)} &
        \(\boldsymbol{\tilde{d}_{H,\mathrm{median}}}\) &
        \textbf{\( \text{IQR}\)} \\
        \midrule
        \rowcolor{rowgray} DNN & 94.8 & 100.0 & 100.0 & 4.0 & 4.0 \\
        AT-rFGSM\(^k\) & 51.4 & 92.0 & 99.5 & 10.0 & 20.0 \\
        \rowcolor{rowgray} AT-MaxMA & \textbf{23.3} & \textbf{65.2} & 96.6 & \textbf{30.0} & 52.0 \\
        KDE & 93.1 & 100.0 & 100.0 & 4.0 & 5.0 \\
        \rowcolor{rowgray} DLA & 91.2 & 100.0 & 100.0 & 4.0 & 4.0 \\
        DNN\textsuperscript{+} & 97.5 & 100.0 & 100.0 & 4.0 & 4.0 \\
        \rowcolor{rowgray} ICNN & 94.8 & 100.0 & 100.0 & 4.0 & 3.0 \\
        PAD-SMA & 36.3 & 89.8 & \textbf{94.6} & 13.0 & 21.0 \\
        \bottomrule
    \end{tabular}
\end{table}

\textbf{Answer to RQ4:} Adaptive attacks expose varying degrees of robustness among defenses. DNN, KDE, DLA, DNN\textsuperscript{+}, and ICNN exhibit substantial vulnerability, with \( \text{ASR}_{10} \) exceeding 90\%. While AT-rFGSM\(^k\) outperforms these simpler defenses, it falls short compared to AT-MaxMA and PAD-SMA. \mbox{AT-MaxMA} demonstrates exceptional robustness at low and moderate perturbation budgets, achieving \( \text{ASR}_{10} = 23.3\%\) and \mbox{\( \text{ASR}_{50} = 65.2\%\).} Conversely, PAD-SMA excels against high-budget attacks, achieving the lowest \( \text{ASR}_{\infty}\) (94.6\%). These results underscore the trade-offs between robustness across different perturbation levels and false positive rates, emphasizing the importance of aligning defenses with specific threat models.

\subsection{RQ5: Trade-offs Between Robustness, Accuracy, and Computational Cost}
\label{subsec:tradeoffs}

To ensure a fair comparison across all defenses, we evaluate only the \textbf{conservative configurations}, in which every input is classified without sample deferral or exclusion. Deferred configurations are omitted from this analysis, as they can inflate performance by discarding ambiguous samples. This setup enables a consistent comparison between detector-based defenses (e.g., KDE, DLA, ICNN, PAD-SMA) and adversarially trained defenses (e.g., AT-rFGSM\(^k\), AT-MaxMA), all operating under identical conditions where each sample must be processed by the classifier.  
The computational cost figures reported in Table~\ref{tab:tradeoff_summary} are derived directly from the normalized training times presented in Appendix~\ref{app:training_cost} (Table~\ref{tab:training_cost_normalized}), ensuring the consistency across all evaluations.

\begin{table*}[tbp]
    \centering
    \caption{Trade-offs among robustness, non-adversarial accuracy, and computational cost for conservative configurations.}
    \label{tab:tradeoff_summary}
    \setlength{\tabcolsep}{6pt}
    \definecolor{rowgray}{gray}{0.95}
    \footnotesize
    \begin{tabular}{lcccc}
        \toprule
        \textbf{Defense} & \textbf{Non-adv. Acc. (\%)} & \textbf{Robustness (\( \text{ASR}_{10}/\text{ASR}_{\infty} \))} & \textbf{Training Cost (×DNN)} & \textbf{Trade-off Summary} \\
        \midrule
        \rowcolor{rowgray} DNN & 97.4 & 94.8 / 100.0 & 1.0 & High accuracy, low robustness, minimal cost \\
        AT-rFGSM\(^k\) & 96.2 & 51.4 / 99.5 & 2.8 & Moderate robustness, efficient training \\
        \rowcolor{rowgray} AT-MaxMA & 94.9 & \textbf{23.3} / 96.6 & 8.9 & Strong low-budget robustness, moderate cost \\
        KDE & 96.2 & 93.1 / 100.0 & 2.4 & Accurate but weakly robust \\
        \rowcolor{rowgray} DLA & 97.2 & 91.2 / 100.0 & 1.1 & Accurate, lightweight, poor robustness \\
        DNN\textsuperscript{+} & 96.2 & 97.5 / 100.0 & 1.1 & Balanced accuracy, negligible robustness gain \\
        \rowcolor{rowgray} ICNN & 93.7 & 94.8 / 100.0 & 1.8 & Low robustness, moderate cost \\
        PAD-SMA & 94.4 & 36.3 / \textbf{94.6} & \textbf{74.8} & Best high-budget robustness, extreme cost \\
        \bottomrule
    \end{tabular}
\end{table*}

\vspace{0.5em}
\noindent\textbf{1) Accuracy Versus Robustness.}
Defenses optimized for adversarial robustness, such as PAD-SMA, AT-rFGSM\(^k\), and AT-MaxMA, exhibit strong resistance to adaptive attacks but show moderate degradation in their classification performance on \textit{non-adversarial samples}.  
As summarized in Table~\ref{tab:tradeoff_summary}, \textbf{PAD-SMA} achieves the highest robustness (\( \text{ASR}_{10}=36.3\%, \text{ASR}_{\infty}=94.6\% \)), yet its non-adversarial accuracy drops to 94.4\%, reflecting the classical robustness–accuracy trade-off~\cite{zhang2019theoretically}.  
A similar pattern is observed for AT-rFGSM\(^k\) and AT-MaxMA, which deliver superior robustness relative to standard models but with a slight loss in clean performance.  
In contrast, defenses with adversary detectors such as KDE, DLA, and ICNN achieve non-adversarial accuracies comparable to adversarially trained models (typically 94–97\%) but collapse under adaptive attacks (\( \text{ASR}_{10} > 90\% \)). 
This indicates that while these models can correctly classify normal samples under clean conditions, they provide little effective robustness once the attacker is aware of and explicitly targets the detector \(g\).

\vspace{0.5em}
\noindent\textbf{2) Robustness versus Computational Cost.}
Training complexity varies significantly across defense categories.  
Lightweight defenses (DNN, DLA, KDE, DNN\textsuperscript{+}, ICNN) complete training within approximately \(1\times\)–\(2\times\) the baseline DNN cost.  
In contrast, adversarially trained defenses incur substantially higher computational overhead due to on-the-fly adversarial example generation.  
\textbf{AT-rFGSM\(^k\)} requires roughly \(2.8\times\) longer training time, while \textbf{AT-MaxMA} increases the cost nearly \(9\times\) owing to its multi-attack generation process.  
The most computationally demanding approach, \textbf{PAD-SMA}, consumes approximately \(75\times\) the baseline training time, as it jointly optimizes a classifier and an adversary detector while synthesizing adversarial samples via a Stepwise Mixture of Attacks.  
Despite this cost, PAD-SMA delivers unmatched robustness under high perturbation budgets, highlighting the inherent cost–robustness trade-off in adversarial training.

\vspace{0.5em}
\noindent\textbf{3) Multi-objective Perspective.}
No single defense simultaneously maximizes robustness, non-adversarial accuracy, and computational efficiency.  
\textbf{AT-MaxMA} achieves the most favorable compromise, combining strong robustness against low- and mid-budget attacks (\( \text{ASR}_{10}=23.3\%, \text{ASR}_{50}=65.2\% \)) with moderate training overhead (\(8.9\times\)).  
\textbf{PAD-SMA} achieves the best robustness at high perturbation levels but at prohibitive cost, making it suitable primarily for high-security environments.  
In contrast, lightweight defenses such as KDE or DLA offer high efficiency and strong performance on non-adversarial data but limited adversarial resistance, fitting low-threat or resource-constrained environments.

\textbf{Answer to RQ5: }The results reveal a clear three-way trade-off among robustness, non-adversarial accuracy, and computational efficiency.  
Adversarially trained defenses (AT-rFGSM\(^k\), AT-MaxMA, PAD-SMA) substantially enhance robustness but incur accuracy loss and higher computational cost.  
Among them, AT-MaxMA achieves the most practical balance, offering robust protection against small perturbations with moderate overhead.  
PAD-SMA attains the highest robustness at large perturbation budgets but at prohibitive cost.  
Detector-based defenses (KDE, DLA, DNN\textsuperscript{+}, ICNN) maintain strong non-adversarial performance and low cost but fail under adaptive attacks.  
These findings underscore the importance of selecting defense strategies based on operational constraints and the anticipated intensity of adversarial threats.

\section{Discussion}
\label{sec:discussion}

The comprehensive evaluation across RQ1–RQ5 provides an integrated perspective on the robustness landscape of binary-constrained malware detection systems. 
The proposed \(\sigma\)-binary attack consistently outperforms existing gradient-based, optimization-based, and ensemble attacks, confirming its capability to effectively operate under strict binary constraints. 
These results demonstrate that robustness in discrete feature spaces remains a fundamental challenge—one that even advanced adversarially trained defenses struggle to overcome.

Adversarially hardened defenses such as AT-MaxMA and PAD-SMA exhibit substantial resilience against low-budget attacks but reveal performance degradation under larger perturbation budgets or adaptive threat models. 
In particular, PAD-SMA demonstrates the highest robustness under strong attacks but incurs elevated false-positive rates and a heavy computational cost, emphasizing that achieving robustness often comes at the expense of practicality. 
Conversely, lighter defenses such as KDE, DLA, and ICNN retain better performance on normal samples but collapse under adaptive attacks.

The results further underscore the inherent three-way trade-off among robustness, accuracy, and efficiency. 
AT-MaxMA achieves the most balanced performance across these dimensions, while PAD-SMA offers superior robustness at the cost of extreme computational overhead. 
These findings emphasize the need for flexible robustness strategies—defenses that can adapt to different operational contexts, security requirements, and computational budgets. 
More broadly, the study reveals that adversarial robustness in binary domains is not solely a model-level property but a system-level characteristic influenced by architectural design, optimization dynamics, and training objectives.

Overall, the empirical insights from this study not only strengthen the understanding of adversarial vulnerabilities in malware detection systems but also provide a methodological foundation for evaluating and improving binary-constrained defenses across other security-critical domains.

\section{Threats to Validity}
\label{sec:validity}

This section outlines the principal validity considerations and methodological limitations of the proposed framework. 
It complements the experimental protocol described in Section~\ref{subsection:experimental_setup} and the ablation analyses in Section~\ref{subsection:sigma-binary attack}.

\textbf{Internal validity.}  
All experiments were performed under a unified setup using fixed random seeds, identical preprocessing pipelines, and consistent software and hardware environments. 
Gradient normalization was applied to mitigate numerical instability. 
While minor non-determinism from GPU operations or floating-point arithmetic may cause negligible variations, their impact on reported results is statistically insignificant.

\textbf{Feature- and problem-space consistency.}  
The proposed attack is designed and evaluated entirely within the \textit{binary feature space}, which serves as the abstraction layer for analyzing adversarial robustness in static malware detectors. 
Within this space, perturbations are explicitly constrained to preserve functional consistency in the underlying problem space. 
To maintain semantic plausibility, we adopt the approximate inverse mapping strategy commonly used in prior studies~\cite{arp2014drebin,li2020adversarial,li2023pad}, where an estimated inverse function~\(\tilde{\phi}^{-1}\) maps binary feature modifications to feasible edits at the application level.
This study follows that established abstraction and does not reimplement or extend the executable-level mapping itself, as such validation lies outside the present research scope. 
Future work should examine direct problem-space realizations to empirically verify functional equivalence of adversarial examples.

\textbf{External validity.}  
Experiments are conducted on the \textit{MalScan} dataset~\cite{wu2019malscan}, selected for its broad temporal coverage (2011–2018) and sufficiently large number of malicious applications (15{,}430 samples), which enables statistically reliable robustness evaluation. 
For comparability with earlier research, the Drebin dataset was initially considered; however, it was excluded because its malware subset (only 5{,}560 samples collected between 2010 and 2012) is both limited in scale and outdated. 
Such a small and temporally narrow corpus restricts the diversity of attack surfaces and hinders meaningful assessment of adversarial robustness, making MalScan a more representative and comprehensive benchmark for this study.
Using MalScan enhances the representativeness of the results for modern malware behaviors. 
Nevertheless, further validation using newer datasets—such as AndroZoo or enterprise-grade corpora—would improve generalizability across evolving threat landscapes.

\textbf{Construct validity.}  
The study employed standard, widely accepted metrics including accuracy, F1 score, false positive and false negative rates, attack success rate (\(\text{ASR}_k\)), median normalized Hamming distance (\(\tilde{d}_{H,\mathrm{median}}\)), and interquartile range (IQR). 
The complete codebase, pretrained models, and hyperparameter configurations are publicly released to ensure reproducibility and facilitate independent verification. 
In addition, the preprocessed version of the MalScan dataset, including standardized feature representations and train/validation/test partitions, is available upon request to support future comparative studies.

\section{Conclusions and Future Work}
\label{sec:conclusion}

This study introduces a robust adversarial attack framework in binary space and a comprehensive evaluation methodology for assessing the robustness of ML-based Android malware detection against adversarial attacks in feature-space. By proposing the \(\sigma\)-binary attack and Prioritized Binary Rounding, we address key limitations in adapting gradient-based attacks—which typically generate continuous adversarial examples—to binary feature domains. Our results emphasize the critical importance of adversarial training in enhancing robustness. However, they also reveal that even advanced defenses, such as PAD-SMA, remain susceptible to adaptive attacks, highlighting vulnerabilities that were previously underestimated. 

This work further identifies inherent trade-offs in robust defenses, including increased false positive rates and computational overhead, underscoring the importance of aligning defense mechanisms with specific operational needs and threat scenarios. These insights provide valuable guidance for designing more effective and efficient defense systems. 

Future research directions could focus on expanding the application of the \(\sigma\)-binary attack and Prioritized Binary Rounding to other binary-constrained domains, such as fraud detection and network intrusion detection, to demonstrate their broader applicability. Another critical direction is the development of standardized benchmarks for evaluating adversarial robustness in binary domains, which would enable consistent and fair comparisons across studies and foster greater transparency in the field.

\bibliographystyle{IEEEtran}
\bibliography{references}

\begin{IEEEbiography}[{\includegraphics[width=1in,clip]{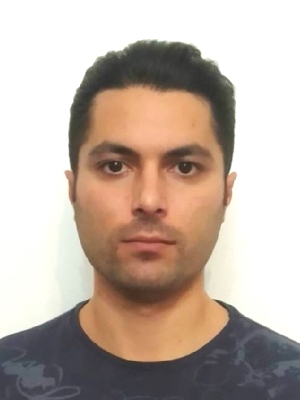}}]{Mostafa Jafari} completed his B.Sc. in Mechanical Engineering from K.N.Toosi University of Technology, followed by his M.Sc. in Software Engineering from Shahid Beheshti University (SBU) in Tehran, Iran. His current research interests revolve around machine learning, adversarial robustness and anomaly detection.
\end{IEEEbiography}

\begin{IEEEbiography}[{\includegraphics[width=1in,clip]{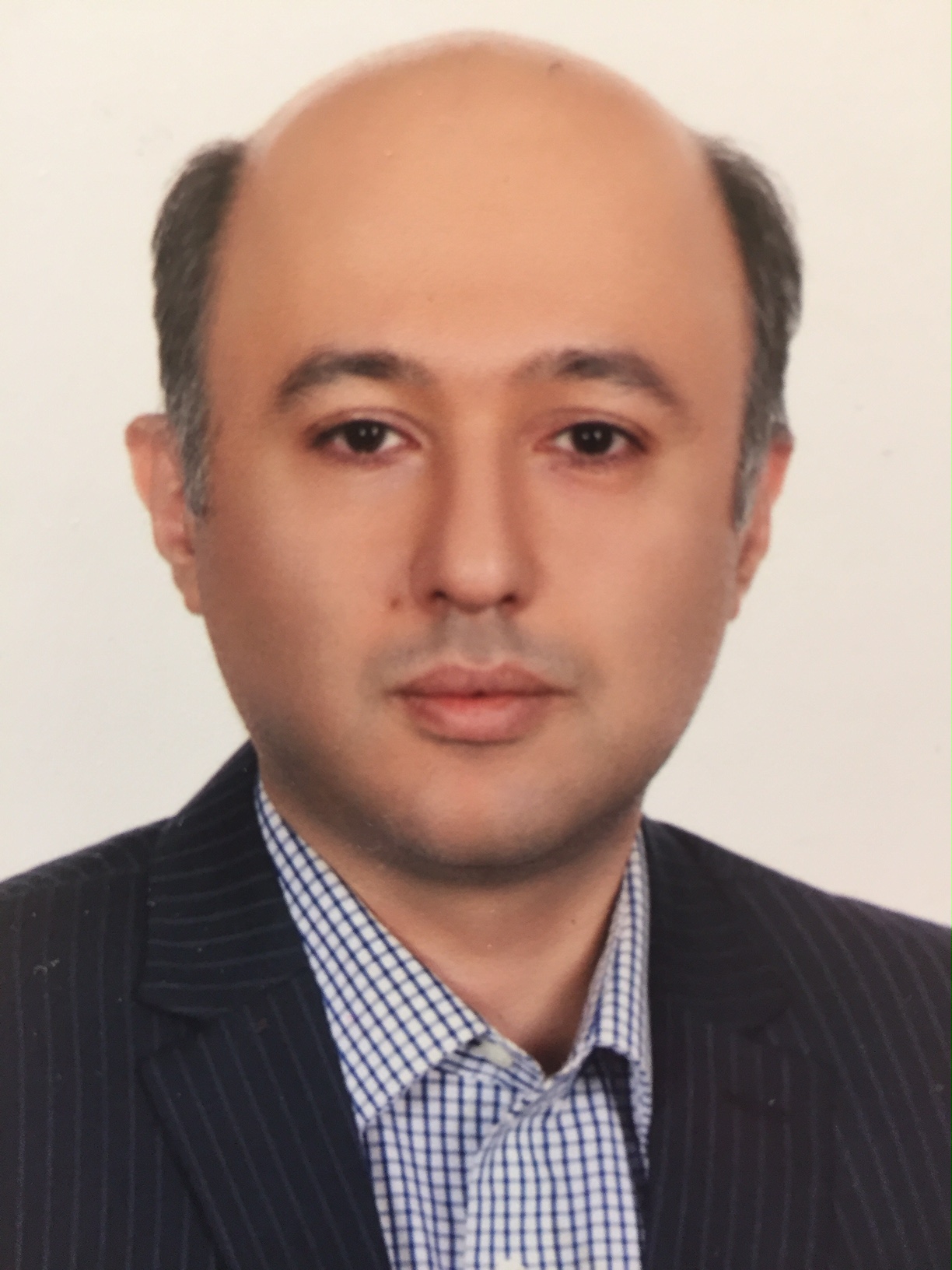}}]{Alireza Shameli-Sendi} is currently an Associate Professor at Shahid Beheshti University. Before joining SBU, he was a Postdoctoral Fellow at Ericsson, Canada and Postdoctoral at ETS and McGill universities in collaboration with Ericsson. He received his Ph.D. degree in computer engineering from Montreal University (Ecole Polytechnique de Montreal), Canada. He obtained his B.Sc. and M.Sc. from Amirkabir University of Technology. His primary research interests include cloud computing and network/information security. He is a recipient of Postdoctoral Research Fellowship Award and Industrial Postdoctoral Fellowship Award from Canada. In addition, he received the best researcher awards, at SBU, in 2018 and 2022.
\end{IEEEbiography}

\clearpage

\appendix
\section{Appendix}

\subsection{Training Cost}
\label{app:training_cost}

Table~\ref{tab:training_cost_normalized} summarizes the training time and relative computational cost of the evaluated defenses described in Section~\ref{subsubsection:defenses}. All experiments were executed under identical hardware and software settings, as detailed in Section~\ref{subsection:experimental_setup}, to ensure fair cross-model comparison.

\begin{table}[h!]
\centering
\caption{Training cost comparison of defenses.}
\label{tab:training_cost_normalized}

\definecolor{rowgray}{gray}{0.95}
\begin{tabular}{lcc}
\toprule
\textbf{Model} & \textbf{Training Time (min)} & \textbf{Relative Cost} \\
\midrule
\rowcolor{rowgray} DNN & 37 & 1.00 \\
AT-rFGSM\(^k\) & 103 & 2.78 \\
\rowcolor{rowgray} AT-MaxMA & 331 & 8.95 \\
KDE & 87 & 2.35 \\
\rowcolor{rowgray} DLA & 41 & 1.11 \\
DNN\textsuperscript{+} & 42 & 1.14 \\
\rowcolor{rowgray} ICNN & 67 & 1.81 \\
PAD-SMA & 2768 & 74.81 \\
\bottomrule
\end{tabular}
\end{table}

As expected, adversarial training substantially increases the overall computational burden compared to standard supervised training. The baseline \textbf{DNN} model, which does not involve adversarial example generation, exhibits the lowest training cost, completing in approximately 37 minutes. In contrast, adversarially trained defenses such as \textbf{AT-rFGSM\(^k\)} and \textbf{AT-MaxMA} require significantly longer training times—roughly \(2.8\times\) and \(9\times\) that of the baseline, respectively. This increase stems from the iterative generation of adversarial perturbations for each mini-batch during training. The multi-attack strategy adopted by AT-MaxMA further amplifies computational demand, as multiple adversarial variants must be synthesized per optimization step.

Hybrid defenses that integrate adversarial detection components (e.g., \textbf{DNN\(^+\)}, \textbf{KDE}, \textbf{DLA}, and \textbf{ICNN}) incur moderate overhead relative to the baseline, primarily due to the additional detector training. These models typically complete training within \(1.1\times\)–\(1.8\times\) the cost of standard DNN training, which reflects the relatively lightweight nature of their secondary detection mechanisms.

The most computationally expensive defense is \textbf{PAD-SMA}, which combines adversarial training with an ICNN-based adversary detector and employs the Stepwise Mixture of Attacks (SMA) for generating diverse adversarial samples. This dual-component training pipeline—requiring simultaneous optimization of the malware detector, the adversary detector, and multiple attack objectives—resulting in a training cost nearly \(75\times\) higher than that of the baseline DNN. Despite this overhead, PAD-SMA achieves the highest overall robustness, illustrating the well-known trade-off between security strength and computational efficiency in adversarial learning.

\subsection{Additional Evaluation under Remaining Defense Models}
\label{app:remaining_defenses}

To complete the evaluation of the proposed \(\sigma\)-binary attack, we extend our analysis to four additional defense mechanisms—KDE, DLA, DNN\(^+\), and ICNN—whose results are reported in Table~\ref{tab:asr_under_defense2} and visualized in Figure~\ref{fig:asr_vs_k2}.  
All experiments were conducted under the same configuration as described in Section~\ref{subsection:experimental_setup}, using the full MalScan test set for reproducibility.

\subsubsection*{Overall Performance}
Across all four defenses, \(\sigma\)-binary consistently achieves the highest attack success rate (\(\text{ASR}_k\)) for all perturbation budgets \(k\), demonstrating near-perfect evasion even under strong detection constraints.  
Specifically, it reaches \(\text{ASR}_{10}=93.8\%\) and \(\text{ASR}_{\infty}=100\%\) against ICNN, and maintains \(\text{ASR}_{10}=91.8\%\) with complete convergence on KDE.  
Under DNN\(^+\) and DLA, \(\sigma\)-binary achieves full success (\(\text{ASR}_{\infty}=100\%\)) across all budgets, indicating robust transferability and adaptability to distinct defense architectures.  
This trend reaffirms the method’s strong generalization beyond adversarially trained defenses.

\subsubsection*{Comparative Analysis}
\begin{figure*}[t!]
    \centering
    \includegraphics[width=\linewidth]{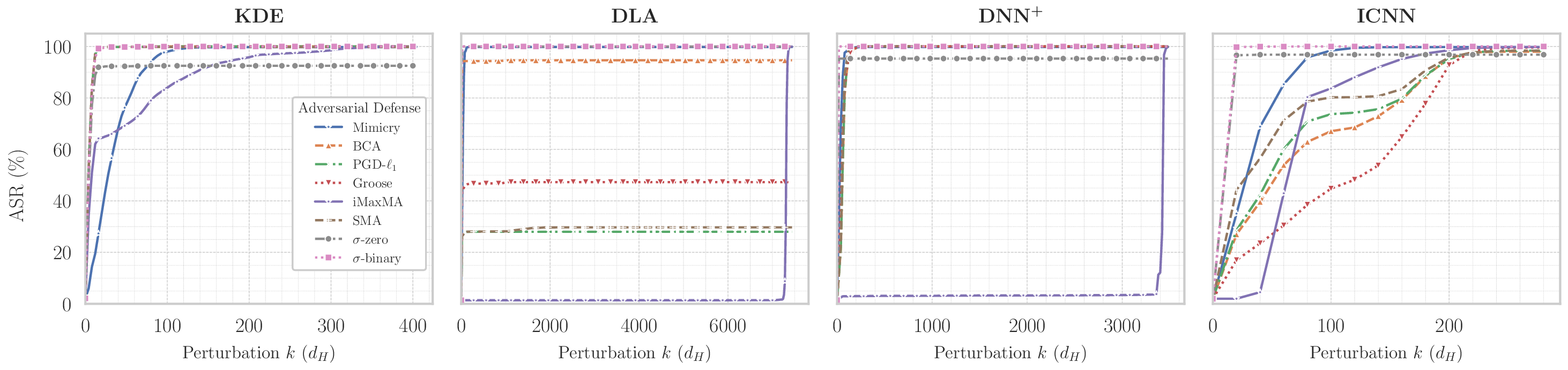}
    \caption{
    Attack success rate (ASR) versus perturbation budget \(k\) (Hamming distance, \(d_H\)) across four defense models (KDE, DLA, DNN\(^+\), and ICNN).  
    Only the most competitive attacks are shown for clarity.}
    \label{fig:asr_vs_k2}
\end{figure*}

The ASR curves in Figure~\ref{fig:asr_vs_k2} reveal clear distinctions between \(\sigma\)-binary and prior attacks.  
While most competing methods (e.g., PGD-\(\ell_p\), SMA, and Mimicry) show early saturation or stagnation at moderate ASR levels, \(\sigma\)-binary exhibits a rapid and monotonic ascent, approaching \(100\%\) at small perturbation budgets.  
Notably, for the KDE and DNN\(^+\) defenses—both highly sensitive to feature smoothness—continuous-space attacks such as CW and PGD-\(\ell_2\) experience severe degradation after binarization, confirming the advantage of binary-aware optimization.  
In contrast, for ICNN and DLA, where feature regularization restricts gradient propagation, \(\sigma\)-binary’s adaptive projection and prioritized rounding enable efficient exploitation of sparse but influential feature subsets, maintaining high ASR even when gradient information is partially constrained.

\subsubsection*{Efficiency Considerations}
In terms of computational efficiency, \(\sigma\)-binary exhibits moderate runtime overheads due to its multi-stage optimization and rounding processes but remains well within practical limits.
For instance, under the ICNN defense, the \(\sigma\)-binary attack records a mean runtime of \(0.30~\text{s/sample}\) with 1.3~GB peak VRAM. 
In contrast, the KDE defense exhibits a higher average runtime of \(2.80~\text{s/sample}\), primarily due to the computational overhead of the kernel-based defense rather than the attack itself. 
Overall, \(\sigma\)-binary maintains a strong balance between attack effectiveness and computational efficiency across all evaluated architectures.

\subsubsection*{Interpretation}
The consistent near-perfect ASR across heterogeneous defenses highlights the robustness and adaptability of \(\sigma\)-binary’s design.  
Its binary-aware optimization framework effectively bridges the gap between gradient-based and discrete feature-optimization strategies, enabling precise, sparse perturbations that remain effective across various model families.  
Moreover, the stability of its ASR curves across DLA, DNN\(^+\), and ICNN demonstrates that its success is not tied to a specific inductive bias or training strategy, reinforcing the general applicability and transferability of the proposed approach.

\definecolor{rowgray}{gray}{0.95}

\begin{table*}[t!]
    \centering
    \caption{Attack Success Rate (ASR, \%) under four defense models (DLA, DNN\(^+\), ICNN, and KDE). 
    Columns report ASR for 10, 20, 50, 100, and $\infty$, along with mean runtime per sample $\bar{t}$ (in seconds/sample) and maximum VRAM usage (in GB). 
    The best result per column is in \textbf{bold}.}
    \vspace{2mm}
    \setlength{\tabcolsep}{4pt}
    \renewcommand{\arraystretch}{1.1}
    \small

    \begin{tabular}{l|ccccccc|ccccccc}
        \toprule
        \multirow{2}{*}{\textbf{Attack}} &
        \multicolumn{7}{c|}{\textbf{DLA}} &
        \multicolumn{7}{c}{\textbf{DNN\textsuperscript{+}}} \\
        \cmidrule(lr){2-15}
         & ASR$_{10}$ & ASR$_{20}$ & ASR$_{50}$ & ASR$_{100}$ & ASR$_{\infty}$ & $\bar{t}$ & VRAM
         & ASR$_{10}$ & ASR$_{20}$ & ASR$_{50}$ & ASR$_{100}$ & ASR$_{\infty}$ & $\bar{t}$ & VRAM \\
        \midrule
        
        \rowcolor{rowgray} BCA & 85.2 & 94.3 & 94.3 & 94.3 & 94.6 & 0.052 & \textbf{0.9} &
        10.0 & 10.9 & 31.3 & 93.7 & \textbf{100.0} & 0.050 & \textbf{0.9} \\
        
        BGA & 1.4 & 1.4 & 1.4 & 1.4 & \textbf{100.0} & 0.045 & 1.1 &
        1.5 & 1.5 & 1.5 & 1.5 & 99.4 & 0.060 & 1.1 \\
        
        \rowcolor{rowgray} Grosse & 44.4 & 44.4 & 44.6 & 46.1 & 47.3 & 0.053 & \textbf{0.9} &
        13.7 & 14.7 & 53.0 & 90.5 & \textbf{100.0} & 0.050 & \textbf{0.9} \\
        
        rFGSM & 1.4 & 1.4 & 1.4 & 1.4 & 99.9 & \textbf{0.001} & 1.0 &
        1.5 & 1.5 & 1.5 & 1.5 & 87.0 & 0.048 & 1.0 \\
        
        \rowcolor{rowgray} PGD-$\ell_1$ & 26.7 & 26.7 & 26.8 & 28.0 & 28.0 & 0.077 & 1.0 &
        10.1 & 11.0 & 33.2 & 94.7 & \textbf{100.0} & 0.071 & 1.0 \\
        
        PGD-$\ell_2$ & 1.4 & 1.4 & 1.4 & 1.4 & \textbf{100.0} & 0.055 & \textbf{0.9} &
        1.5 & 1.5 & 1.5 & 1.5 & 86.5 & 0.052 & \textbf{0.9} \\
        
        \rowcolor{rowgray} PGD-$\ell_\infty$ & 1.4 & 1.4 & 1.4 & 1.4 & \textbf{100.0} & 0.056 & 0.9 &
        1.5 & 1.5 & 1.5 & 1.5 & 86.5 & 0.054 & \textbf{0.9} \\
        
        iMaxMA & 1.4 & 1.4 & 1.4 & 1.4 & \textbf{100.0} & 0.180 & 1.1 &
        1.71 & 2.3 & 2.5 & 2.7 & \textbf{100.0} & 0.171 & 1.1 \\
        
        \rowcolor{rowgray} SMA & 26.5 & 26.5 & 26.6 & 28.0 & 29.0 & 0.239 & 1.1 &
        12.6 & 14.3 & 43.1 & 98.8 & \textbf{100.0} & 0.109 & 1.1 \\

        Mimicry & 14.7 & 33.9 & 78.7 & 99.3 & 99.8 & 0.012 & 1.2 &
        17.1 & 34.7 & 79.2 & 99.1 & 99.8 & \textbf{0.012} & 1.3 \\
        
        \rowcolor{rowgray} $\sigma$-zero & 89.3 & 99.3 & 99.4 & 99.4 & 99.4 & 0.114 & 1.1 & 
        96.1 & 99.7 & 99.7 & 99.7 & 99.7 & 0.113 & 1.1 \\
        
        CW & 1.4 & 1.4 & 1.4 & 1.4 & 8.0 & 0.444 & 2.2 &
        10.2 & 10.2 & 10.2 & 10.2 & 10.2 & 0.433 & 2.2 \\
        
        \rowcolor{rowgray} $\sigma$-binary & \textbf{90.0} & \textbf{100.0} & \textbf{100.0} & \textbf{100.0} & \textbf{100.0} & 0.222 & 1.3 &
        \textbf{97.2} & \textbf{100.0} & \textbf{100.0} & \textbf{100.0} & \textbf{100.0} & 0.144 & 1.3 \\
        \bottomrule
    \end{tabular}

    \vspace{3mm} 

    \begin{tabular}{l|ccccccc|ccccccc}
        \toprule
        \multirow{2}{*}{\textbf{Attack}} &
        \multicolumn{7}{c|}{\textbf{ICNN}} &
        \multicolumn{7}{c}{\textbf{KDE}} \\
        \cmidrule(lr){2-15}
         & ASR$_{10}$ & ASR$_{20}$ & ASR$_{50}$ & ASR$_{100}$ & ASR$_{\infty}$ & $\bar{t}$ & VRAM
         & ASR$_{10}$ & ASR$_{20}$ & ASR$_{50}$ & ASR$_{100}$ & ASR$_{\infty}$ & $\bar{t}$ & VRAM \\
        \midrule
        
        \rowcolor{rowgray} BCA & 22.2 & 27.0 & 46.8 & 67.0 & 98.0 & 0.078 & \textbf{0.9} &
        91.6 & 99.6 & 99.8 & \textbf{100.0} & \textbf{100.0} & 0.526 & 2.6 \\
        
        BGA & 2.1 & 4.0 & 9.5 & 19.7 & 48.5 & 0.085 & 1.1 &
        2.1 & 2.1 & 2.1 & 2.1 & 2.1 & 1.102 & 2.6 \\
        
        \rowcolor{rowgray} Grosse & 14.0 & 17.0 & 26.7 & 44.8 & 99.6 & 0.076 & \textbf{0.9} &
        91.6 & 99.6 & 99.8 & \textbf{100.0} & \textbf{100.0} & \textbf{0.050} & 2.6 \\
        
        rFGSM & 2.0 & 2.0 & 2.0 & 2.0 & 97.3 & 0.075 & 1.0 &
        2.2 & 2.9 & 25.9 & 80.9 & 91.1 & 0.666 & \textbf{2.2} \\
        
        \rowcolor{rowgray} PGD-$\ell_1$ & 23.9 & 28.6 & 52.0 & 73.7 & 98.4 & 0.098 & 1.0 &
        91.6 & 99.6 & 99.8 & \textbf{100.0} & \textbf{100.0} & 0.519 & 2.3 \\
        
        PGD-$\ell_2$ & 2.0 & 2.0 & 2.0 & 2.1 & 98.6 & 0.078 & 1.0 &
        3.0 & 3.7 & 10.8 & 34.6 & 64.5 & 1.082 & 2.6 \\
        
        \rowcolor{rowgray} PGD-$\ell_\infty$ & 2.0 & 2.0 & 6.4 & 57.2 & 98.9 & 0.080 & 1.0 &
        2.1 & 2.1 & 2.1 & 2.4 & 54.3 & 1.082 & 2.6 \\
        
        iMaxMA & 2.0 & 2.0 & 2.0 & 2.8 & 99.8 & 0.247 & 1.1 &
        37.5 & 45.1 & 79.1 & 61.2 & 100.0 & 2.653 & 3.1 \\
        
        \rowcolor{rowgray} SMA & 38.4 & 43.8 & 63.5 & 80.2 & 98.4 & 0.099 & 1.1 &
        91.7 & 99.6 & 99.7 & 99.8 & 99.8 & 0.839 & 2.9 \\

        Mimicry & 16.6 & 35.1 & 77.8 & 98.4 & 99.8 & \textbf{0.019} & 1.3 &
        16.5 & 35.0 & 77.7 & 97.9 & 99.7 & 2.987 & 2.9 \\
        
        \rowcolor{rowgray} $\sigma$-zero & 92.8 & 98.8 & 99.0 & 99.1 & 99.1 & 0.151 & 1.2 & 
        82.7 & 90.6 & 90.6 & 90.8 & 90.8 & 1.148 & 2.6 \\
        
        CW & 2.4 & 2.4 & 2.4 & 2.4 & 2.4 & 0.698 & 2.3 &
        2.1 & 2.2 & 2.2 & 2.2 & 2.2 & 8.901 & 2.7 \\
        
        \rowcolor{rowgray} $\sigma$-binary & \textbf{93.8} & \textbf{99.8} & \textbf{100.0} & \textbf{100.0} & \textbf{100.0} & 0.304 & 1.3 &
        \textbf{91.8} & \textbf{99.8} & \textbf{99.9} & \textbf{100.0} & \textbf{100.0} & 2.799 & 2.5 \\
        \bottomrule
    \end{tabular}

    \label{tab:asr_under_defense2}
\end{table*}

\end{document}